%% file: ms.tex
\documentclass[a4paper,fleqn,usenatbib]{mnras}
\usepackage{footmisc}
\usepackage{graphicx}
\usepackage{amssymb}
\usepackage{amsmath}
\usepackage{wasysym}
\usepackage{color}
\usepackage[normalem]{ulem}
\usepackage{psfrag}
\usepackage{multirow,bigdelim}
\usepackage{pdflscape}
\graphicspath{{./}{../plots/}}

\newcommand{\snia}{\hbox{SN~Ia}}
\newcommand{\sneia}{\hbox{SNe~Ia}}

\newcommand{\snoopy}{\hbox{SNooPy}}

\newcommand{\mstar}{\hbox{$M_{\star}$}}

\newcommand{\msun}{\hbox{M$_{\odot}$}}

\newcounter{minirefcount}
\title[Relative SN Ia Rates From ASAS-SN]{The Relative Specific Type Ia Supernovae Rate From Three Years of ASAS-SN}

\author[J. S. Brown et al.]{J. S. Brown,$^{1}$\thanks{E-mail: brown@astronomy.ohio-state.edu}
K. Z. Stanek,$^{1,2}$
T. W.-S. Holoien,$^{3}$
C. S. Kochanek,$^{1,2}$\newauthor
B. J. Shappee,$^{4}$
J. L. Prieto,$^{5,6}$
S. Dong,$^{7}$
P. Chen,$^{7}$
Todd. A. Thompson,$^{1,2}$\newauthor
J. F. Beacom,$^{1,2,8}$
M. D. Stritzinger,$^{9}$
D. Bersier,$^{10}$
and J. Brimacombe$^{11}$\\
$^{1}$ Department of Astronomy, The Ohio State University, 140 West 18th Avenue, Columbus, OH 43210, USA\\
$^{2}$ Center for Cosmology and Astro-Particle Physics, The Ohio State University, 191 West Woodruff Avenue, Columbus, OH 43210, USA\\
$^{3}$ Carnegie Observatories, 813 Santa Barbara Street, Pasadena, CA 91101, USA\\
$^{4}$ Institute for Astronomy, University of Hawai’i, 2680 Woodlawn Drive, Honolulu, HI 96822\\
$^{5}$ N\'ucleo de Astronom\'ia de la Facultad de Ingenier\'ia y Ciencias, Universidad Diego Portales, Av. Ej\'ercito 441, Santiago, Chile \\
$^{6}$ Millennium Institute of Astrophysics, Santiago, Chile \\
$^{7}$ Kavli Institute for Astronomy and Astrophysics, Peking University, Yi He Yuan Road 5, Hai Dian District, Beijing 100871, China\\
$^{8}$ Department of Physics, The Ohio State University, 191 W. Woodruff Ave, Columbus, OH 43210, USA\\
$^{9}$ Department of Physics and Astronomy, Aarhus University, Ny Munkegade 120, DK-8000 Aarhus C, Denmark\\
$^{10}$ Astrophysics Research Institute, Liverpool John Moores University, 146 Brownlow Hill, Liverpool L3 5RF, UK\\
$^{11}$ Coral Towers Observatory, Cairns, QLD 4870, Australia\\
}

\date{Accepted XXX. Received YYY; in original form ZZZ}

\pubyear{2017}

\begin{document}
\label{firstpage}
\pagerange{\pageref{firstpage}--\pageref{lastpage}}
\maketitle

\begin{abstract}
We analyze the 476 SN Ia host galaxies from the All-Sky Automated Survey for Supernova (ASAS-SN) Bright Supernova Catalogs to determine the observed relative Type Ia supernova (SN) rates as a function of luminosity and host galaxy properties. We find that the luminosity distribution of the \sneia\ in our sample is reasonably well described by a Schechter function with a faint-end slope $\alpha \approx 1.5$ and a knee $M_{\star} \approx -18.0$. Our specific  SN Ia rates are consistent with previous results but extend to far lower host galaxy masses. We find an overall rate that scales as $(\mstar/10^{10}\msun)^{\alpha}$ with $\alpha \approx -0.5$. This shows that the specific SN Ia rate continues rising towards lower masses even in galaxies as small as $\log(M_{\star}/M_{\odot}) \lesssim 7.0$, where it is enhanced by a factor of $\sim10-20$ relative to host galaxies with stellar masses $\sim10^{10}\msun$. We find no strong dependence of the specific \snia\ rate on the star formation activity of the host galaxies, but additional observations are required to improve the constraints on the star formation rates.
\end{abstract}


\section{Introduction}
\label{sec:intro}

Type Ia supernovae (\sneia), which arise from the thermonuclear detonation of carbon-oxygen white dwarfs \citep{Hoyle60}, are a fundamental pillar of modern astronomy, cosmology, and physics. These events are unambiguously classified with low resolution optical spectra \citep{Filippenko97}, and evolve in such a way that their intrinsic luminosities and thus distances can be inferred with relatively high precision \citep{Phillips93,Hamuy95,Riess96}. The widespread interest in \sneia\ has been primarily driven by their luminosity and homogeneity, which makes them excellent probes of the large scale universe and cosmic evolution \citep[e.g.,][]{Riess98,Perlmutter99}. Given the pivotal role of \sneia\ in our understanding of the fundamental constants of our universe, and the tension with other independent cosmological experiments \citep[e.g.,][]{Planck16}, it is paramount that we expand our understanding of their origins. 

Unfortunately, our picture of \sneia\ is not as constrained as one might hope. Even the physical systems that give rise to the explosions are not well characterized \citep[for reviews, see][]{MaozMannucci12,Wang12}. The two competing theories both involve a carbon-oxygen WD in a close binary. In the single-degenerate (SD) scenario \citep{Whelan73,Nomoto82}, the binary companion is a non-degenerate star which steadily transfers mass onto the WD until a thermonuclear runaway occurs. In the double-degenerate (DD) scenario \citep{Tutukov79,Iben84,Webbink84}, a merger or collision \citep[e.g.,][]{Thompson11, Dong15} of two white dwarfs provides the necessary conditions for explosive burning of the carbon-oxygen fuel. Observational evidence disfavors the presence of a SD companion in the most well-studied cases \citep{Nugent11,Chomiuk12,Shappee13,Shappee18}. On the other hand, there are theoretical difficulties with producing a \snia\ from the DD scenario \citep[e.g.,][]{Shen12}.

One avenue for progress is the characterization of the delay-time distribution (DTD) of \sneia, or the \snia\ rate as a function of time after an episode of star formation. By constraining the rate at which \sneia\ occur after an episode of star formation, certain progenitor scenarios can be ruled out. The \snia\ DTD is broadly consistent with a $t^{-1}$ form; equivalently, there is evidence for a population of \sneia\ that occur promptly after star formation ($t\sim10^8$~yr), and a delayed component that occurs at much later times ($t\gtrsim10^9$~yr) \citep{Mannucci05, Scannapieco05, Sullivan06, Brandt10, Maoz11, Maoz12}.

An alternative approach is characterizing the \snia\ host galaxy population. Observationally, lower mass galaxies produce more \sneia\ per unit stellar mass than high-mass galaxies \citep[e.g.,][]{Mannucci05}. This has motivated several studies geared towards improving our understanding of the relationship between SNe Ia and their host galaxies. In particular, the Lick Observatory Supernova Search \citep[LOSS;][]{Li00}, the Nearby Supernova Factory \citep[SNfactory;][]{Aldering02,Childress13a}, the Texas Supernova Search \citep[TSS;][]{Quimby06}, the SuperNova Legacy Survey \citep[SNLS;][]{Astier06,Guy10}, the Sloan Digital Sky Survey--II Supernova Survey \citep{Frieman08}, and the Palomar Transient Facility \citep[PTF;][]{Law09} identified several trends between SNe Ia properties and their host galaxies \citep[e.g.,][]{Neill09, Sullivan10, Lampeitl10, Pan14}, as well as the relative SN Ia rate as a function of host galaxy properties \citep[e.g.,][]{Neill06, Sullivan06, Quimby12, Smith12, Gao13, Loss3, Graur13, Graur15, Graur17, Heringer17}.

A more contentious issue is whether these trends extend to local environments within the host galaxies. Characterizing this relationship is important, since the residuals of fits to the dependence of distance on redshift (i.e., Hubble residuals) are correlated with host galaxy properties \citep[e.g.,][]{Lampeitl10, Sullivan10, Kelly10, Gupta11, Johansson13, Childress13b, Pan14, Wolf16, Uddin17}. Several studies have suggested that Hubble residuals are indeed correlated with local environment \citep{Rigault13, Rigault15, MorenoRaya16, Roman17}, while \citet{Jones15} argue that there is no dependence on local star formation rate (SFR). Similarly, \citet{Anderson15} used an independent sample of SNe and recovered the dependence of \snia\ properties on host galaxy parameters, but found no dependence of \snia\ properties on the local environment. Of course, the general galaxy population evolves with redshift \citep{Madau96,Hopkins06,Behroozi13,Madau14}, which makes understanding these trends critical for utilizing high redshift SNe in cosmological studies.

The strategies employed in most SN surveys suffer from observational biases and incompleteness problems that the All-Sky Automated Survey for Supernovae \citep[ASAS-SN;][]{Shappee13,Kochanek17} was designed to minimize. ASAS-SN monitors the entire night sky at a relatively high cadence. The discovery and rapid propagation of nearby, bright transients to the astronomical community allows for detailed follow-up with both ground and space based instrumentation. ASAS-SN has been influential in the discovery of a wide variety of transients including novel SNe \citep[e.g.,][]{Dong16, Godoy17, Holoien16_acta,Shappee16,Shappee18_18bt, Bose18, Vallely18}, tidal disruption events \citep[TDEs;][]{Holoien14,Holoien16_14li,Holoien16_15oi, Brown16_14ae, Brown16_14li}, flares in active galactic nuclei \citep[AGN;][]{Shappee14}, stellar outbursts \citep{Holoien14, Schmidt14, Schmidt16, Herczeg16}, and cataclysmic variable stars \citep[CVs;][]{Kato14a, Kato14b, Kato15, Kato16}. Additionally, ASAS-SN data has played a crucial role in constraining the pre-discovery and early-time light curves of several other interesting objects \citep[e.g.,][]{Bose17}.

While the discovery and follow-up of these rare objects is informative, the statistical power of ASAS-SN has yet to be exploited. ASAS-SN is largely agnostic with regard to host galaxy properties and thus provides a quasi-unbiased census of SNe in the nearby universe. Some SNe are invariably missed due to their location on the sky, being near bright stars or behind the Sun, and extinction (both Galactic and extragalactic) will also result in some incompleteness. However, for the optically accessible, bright SNe ($m_V < 17$), ASAS-SN is more sensitive to small galaxies and nuclear regions than most previous SN surveys \citep{snCat2,snCat3,snCat1}. Furthermore, the brightness and sample size of the ASAS-SN survey has allowed us to spectroscopically follow-up and classify all of our discoveries, which eliminates a significant source of uncertainty and bias that has affected many previous SN surveys.

In this paper we perform a census of SN Ia host galaxies using data from the first $\sim3$ years of ASAS-SN. In Section~\ref{sec:data}, we describe the SN Ia sample and the archival data used to analyze the host galaxies. In Section~\ref{sec:analysis}, we analyze the mass distribution of the SN Ia host galaxies and derive the observed specific SN Ia rate for an unprecedentedly wide range of stellar masses. In Section~\ref{sec:conclusions}, we summarize our findings and discuss future directions.

\section{Data}
\label{sec:data}

\subsection{The SN Ia Sample}

\begin{figure}
\centering{\includegraphics[scale=1.,width=0.49\textwidth,trim=0.pt 0.pt 0.pt 0.pt,clip]{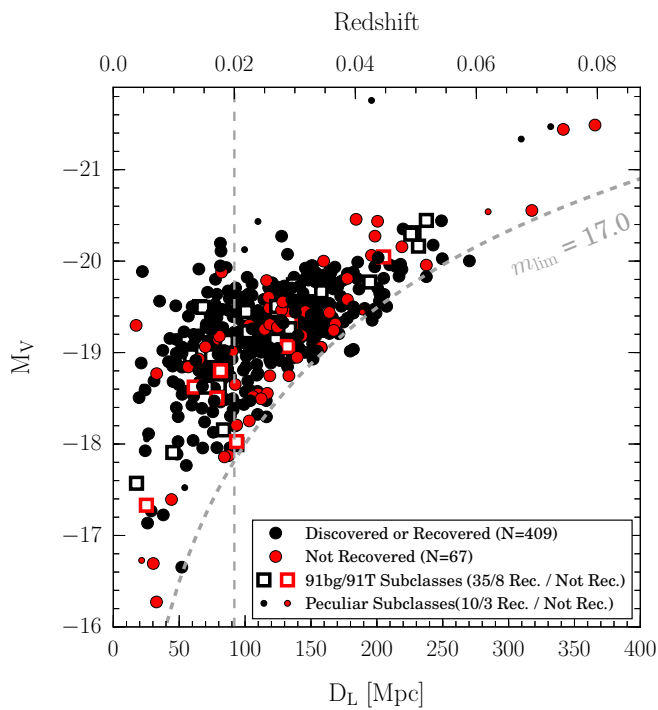}}
\caption{SN Ia peak absolute magnitudes after applying corrections for Galactic reddening and redshift ($K$ correction), versus luminosity distance (redshift is shown on the top axis). The dashed gray curve denotes the detection threshold for a survey with a limiting magnitude $m_V = 17.0$ mag, assuming no extinction along the line of sight. The vertical gray dashed line denotes the distance limit used to construct our volume-limited sample. Black (red) symbols show SNe discovered/recovered (not recovered) by ASAS-SN. Circles represent normal SNe Ia, and squares represent those belonging to the 91bg and 91T subclasses. The small points denote peculiar SNe Ia (including Ia-02cx) that we exclude from our analysis.}
\label{fig:absMag_dist}
\end{figure}

\begin{figure}
\centering{\includegraphics[scale=1.,width=0.49\textwidth,trim=0.pt 0.pt 0.pt 0.pt,clip]{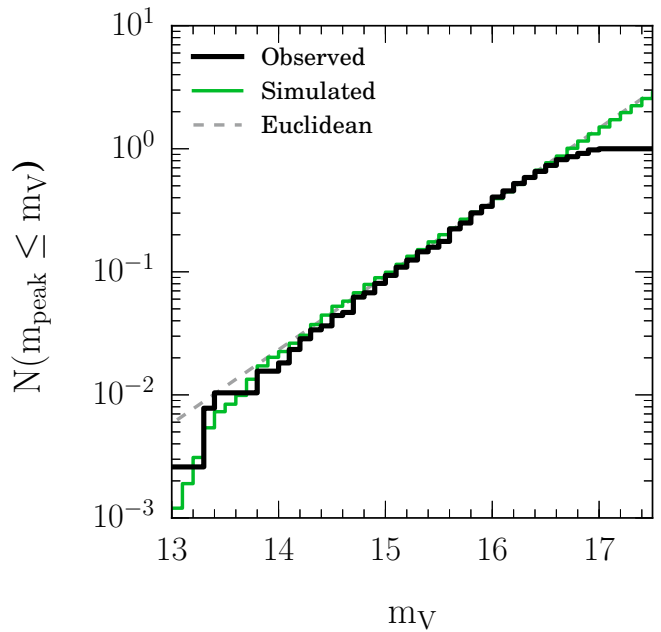}}
\caption{Cumulative number of SNe as a function of apparent magnitude. The black histogram shows the observed distribution, the gray dashed line shows the expected number in a Euclidean universe, and the green histogram shows the results from our simulation. We normalize the Euclidean and simulated curves to the number of observed SNe discovered brighter than $m_V = 16$. Using the differential form of this distribution, we derive the completeness as a function of apparent magnitude by computing the ratio of the number of observed SNe to the number of expected SNe at a given brightness.}
\label{fig:completeness}
\end{figure}

\begin{figure*}
\centering{\includegraphics[scale=1.,width=0.49\textwidth,trim=0.pt 0.pt 0.pt 0.pt,clip]{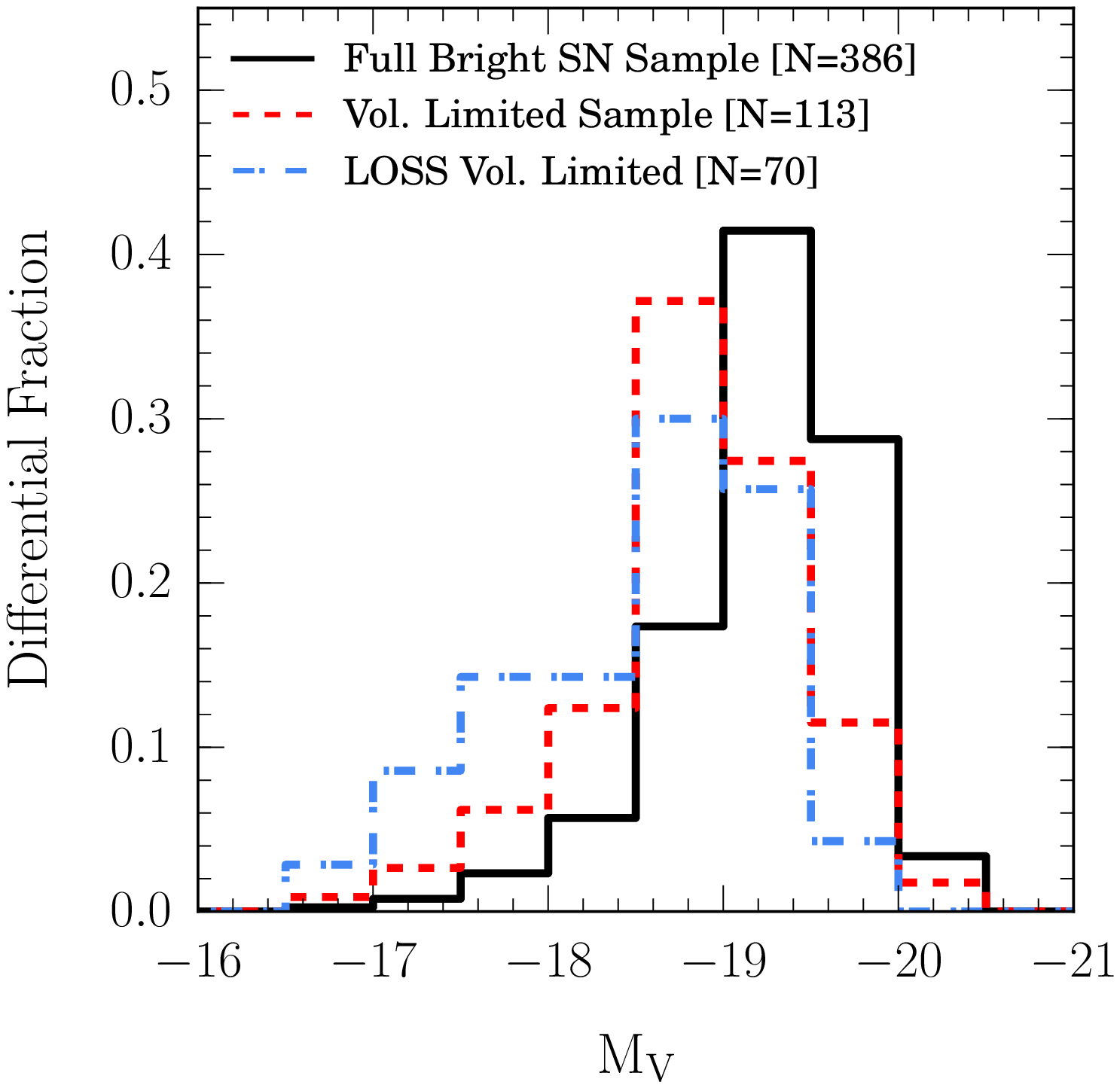}}
\centering{\includegraphics[scale=1.,width=0.49\textwidth,trim=0.pt 0.pt 0.pt 0.pt,clip]{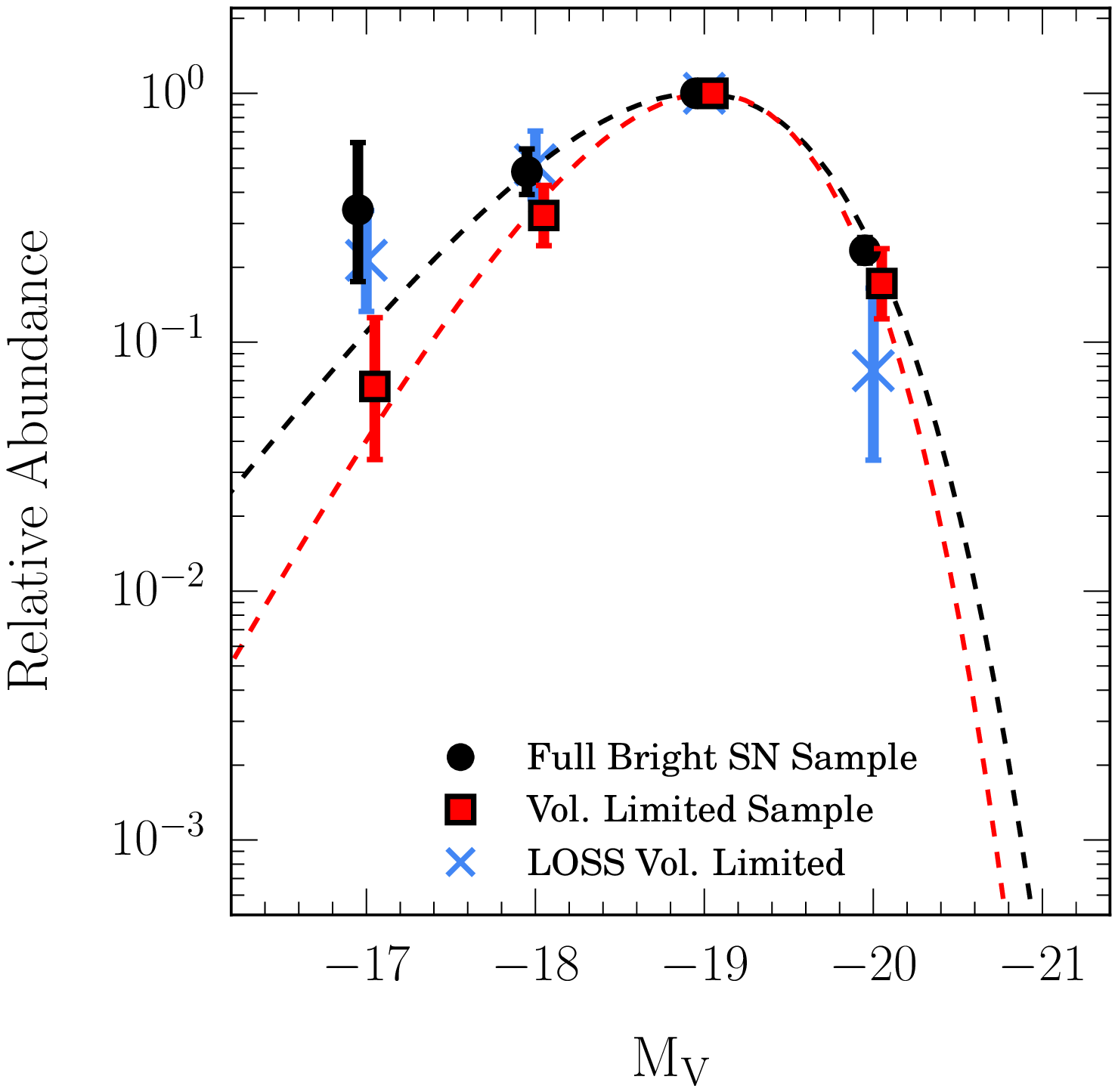}}
\caption{Left: The observed distributions of SN Ia absolute magnitudes. The full and volume-limited samples presented here are shown by the solid black and dashed red histograms, respectively. For comparison, the dot-dashed blue histogram shows the distribution from the volume-limited LOSS sample \citep{Loss2}. The volume-limited samples have a larger fraction of relatively faint SNe than the full sample, as one would expect. Right: True relative luminosity function (i.e., corrected for completeness, redshift, and luminosity bias). The dashed lines show our fits to the \snia\ luminosity function.}
\label{fig:lumFunc}
\end{figure*}

The SN Ia sample is constructed from the ASAS-SN Bright Supernova Catalogs \citep{snCat2,snCat3,snCat1}. These catalogs contain 476 SNe Ia in total. ASAS-SN discovered 325 of these SNe Ia, and out of the 151 not discovered by ASAS-SN, 84 were recovered in ASAS-SN data. To build our unbiased sample of \sneia, we selected \snia\ hosts for which the transient peaked brighter than $m_V = 17$ and was either discovered by ASAS-SN or recovered after discovery in ASAS-SN data. In order to construct the most complete and unbiased sample as possible, we exclude the 67 \sneia\ that were neither discovered nor recovered by ASAS-SN, and the 13 \sneia\ that did not peak brighter than $m_V = 17$. Additionally, we exclude \sneia\ that were classified as peculiar (e.g., Ia-pec or Ia-CSM), Ia-02cx events \citep{Li03}, and other rare and unusual subtypes (Ia-06bt, Ia-07if, Ia-09dc). We retain the Ia-91bg and Ia-91T subtypes in our analysis, since they likely constitute the tails of the distribution of ``normal'' \sneia.

The resulting sample represents a nearly complete and unbiased census of SNe Ia in the nearby universe \citep{snCat3}. However, the sample is not entirely complete. A fraction of SNe are either undiscovered or excluded from the sample due to essentially random effects (e.g., the position of the Sun, high Galactic extinction, or bright foreground sources). These sources of incompleteness are physically unassociated with the SNe, and thus do not bias the calculation of \textit{relative} quantities in any significant way. There is also likely a population of \snia\ that are missed due to extinction in the host galaxy. However, out of all SN types, \sneia\ are the most weakly associated with star-forming regions \citep{Anderson15}, so the immediate local environment of \sneia\ is unlikely to be a significant source of systematic incompleteness. This is also clearly seen in models for the detection of SNe in the Milky Way \citep{Adams13}. There are populations of dusty galaxies in which SNe would be obscured in a significant volume of the host \citep[e.g., ultraluminous infrared galaxies or ULIRGs][]{Lonsdale06}, but at low redshift these galaxies are rare. \citet{Goto11} found that LIRGs and ULIRGs are responsible for $\lesssim10\%$ and $\lesssim0.5\%$ of the total infrared luminosity in the local universe, respectively, which means that these heavily dust obscured galaxies are not representative of the underlying stellar mass distribution in the universe. Furthermore, the SNe in these dusty galaxies are not necessarily missed by modern optical surveys: SN 2014J \citep{Fossey14} was discovered in the optical despite the fact that the host galaxy M82 is relatively dusty by low redshift standards. With these considerations in mind, the ASAS-SN sample is the closest realization of a statistically complete and unbiased supernova survey of the nearby universe to date.

The SN sample constructed here is, by design, a magnitude limited sample. All else being equal, the relatively luminous SNe Ia (and their host galaxies) will be over represented, since the effective survey volume scales roughly as $L_{\rm SN}^{3/2}$. In practice, the situation is more complex. While massive host galaxies host fainter, faster SNe Ia \citep{Hamuy00}, the intrinsic luminosities of SNe Ia of a given color and stretch (i.e., after correction) are higher in more massive galaxies. Furthermore, many of the SNe in our sample are found deep within their host galaxies, where extinction may be non-negligible. In fact, ASAS-SN is more sensitive to SNe embedded within the starlight of big galaxies than other surveys \citep{snCat3}. To minimize the dependence on completeness corrections, we also construct a volume-limited sample ($z<0.02$; $D_L \lesssim 90$~Mpc), which should not be significantly affected by the correlation between SN Ia brightness and host galaxy mass.

In Figure~\ref{fig:absMag_dist} we show the distribution of observed SN Ia absolute magnitudes. We have applied a correction for Galactic reddening \citep{Cardelli89,Odonnell94,Schlafly11} and a $K$ correction \citep{Kim96,Hogg02}. The $K$ corrections are computed with \snoopy\ \citep{Burns11}, which uses the SN Ia templates from \citet{Hsiao07}. The magnitudes have not been corrected for the local reddening from the host galaxy. Thus, the absolute magnitudes are simply

\begin{flalign}
M_V = m_V - A_{V,{\rm Gal}} - \mu(z) - K(z).
\label{eqn:absMag}
\end{flalign}

\noindent We show these absolute magnitudes versus luminosity distance with the redshift shown on the top axis. The black points denote SNe either discovered or recovered by ASAS-SN, while the red points denote SNe that were not recovered by ASAS-SN and are excluded from our analysis. The squares represent SN Ia subtypes, and the small circles represent the peculiar SNe Ia that are excluded from our analysis. The gray dashed curve shows the detection threshold for a magnitude limited survey with a limiting magnitude $m_V = 17.0$. The vertical gray dashed line shows the distance limit for our volume-limited sample. Within this volume, we expect to recover nearly all SN Ia with absolute magnitudes $M_V \lesssim -18$, which is $\gtrsim80\%$ of SNe Ia \citep{Loss2}. For SNe Ia with absolute magnitudes $M_V \gtrsim -18$, even our volume-limited sample is incomplete, but this is a relatively small fraction of SNe.

In Figure~\ref{fig:completeness} we show the relative completeness of our sample as a function of peak apparent magnitude. The black histogram shows the observed cumulative distribution; the dashed gray line shows the expected distribution in a Euclidean universe, normalized to the number of SNe discovered at $m_V<16.0$. In order to gain a better estimate of the expected distribution in $m_V$, we simulate $10^5$ SNe uniformly distributed in comoving volume, with distances up to $500$~Mpc. For each SN, we randomly generate an absolute magnitude drawn from a Gaussian distribution with mean $M_V=-18.5$ mag and $\sigma = 1.0$ mag, which (as we will show) is a reasonable approximation of the \snia\ luminosity function \citep{Loss2}. We weight each SN by a factor of $(1+z)^{-1}$ to account for the time dilation of SN rates. The resulting distribution is shown in Figure~\ref{fig:completeness} by the green histogram. The simulated distribution differs little from the expectation for a Euclidean universe. At $m_V\sim17$, our sample is roughly $\sim70\%$ complete, consistent with the results from \citet{snCat3}. We use the differential form of this distribution in order to compute a completeness correction as a function of apparent magnitude for SNe fainter than $m_V\sim16$. We apply no completeness correction to the bright end of the distribution where the differences are dominated by statistical fluctuations. 

Once we have the completeness corrections, we can estimate the relative luminosity function of SNe Ia, which we show in Figure~\ref{fig:lumFunc}. In the left panel we show the observed distribution of absolute magnitudes. The black histogram shows the full sample, and the red histogram shows the volume-limited sample; the dot dashed blue histogram shows the distribution of the volume-limited LOSS sample \citep{Loss2}. The volume-limited samples have a higher fraction of low luminosity SNe Ia than the full magnitude limited sample. We convert the ASAS-SN distributions into estimates of the true luminosity function using the $V/V_{\rm max}$ method \citep{Schmidt68,Huchra73,Felten76}. For each SN, we compute the maximum volume ($V_{\rm max}$) in which the SN could be recovered by a survey with a limiting magnitude $m_V = 16.8$. This is an empirical limiting magnitude; this value produces a median value of $V/V_{\rm max}$ close to 0.5, which is to be expected if sources uniformly populate the survey volume. We compute the relative luminosity function for each bin in absolute magnitude centered on $M$ as

\begin{equation}
\Phi(M) = \sum_{i=1}^{N} \frac{1}{V_{M,i}} \times w_i \times(1+z_i),
\label{eqn:sum}
\end{equation}

\noindent where the sum is over all the SNe within the bin. The weights $w_i$ correct for the incompleteness given the apparent peak magnitude of each SN, and the factor of $(1+z)$ accounts for time dilation. The results are shown in the right panel of Figure~\ref{fig:lumFunc}. The black circles show the relative luminosity function computed from the full sample, the red squares show the results for the volume-limited sample, and the blue crosses show the control-time weighted counts from \citet{Loss2}. The luminosity functions are normalized to the bin at $M_V = -19$. In this paper we are not aiming for an absolute rate calibration. The shape of the relative luminosity function is consistent with the volume-limited luminosity function presented in \citet{Loss2}. We fit a \citet{Schechter76} function

\begin{equation}
\phi(L) \propto \left(\frac{L}{L_*}\right)^{\alpha} \exp\left(-\frac{L}{L_*}\right)
\label{eqn:sch}
\end{equation}

\noindent to the relative luminosity function of both the full and volume limited samples, where $\alpha$ is the faint-end slope, and $L_*$ (alternatively $M_*$ in magnitude space) determines the ``knee'' of luminosity function. Our fits are shown in Figure~\ref{fig:lumFunc} as dashed lines. We find $(\alpha,M_*)$ corresponding to $(1.3\pm0.4,-18.1\pm0.1)$ and $(2.1\pm0.3,-17.8\pm0.1)$ for the full sample and the volume-limited sample, respectively.

\subsection{Archival Host Data}

\begin{figure*}
\centering{\includegraphics[scale=1.,width=0.49\textwidth,trim=0.pt 0.pt 0.pt 0.pt,clip]{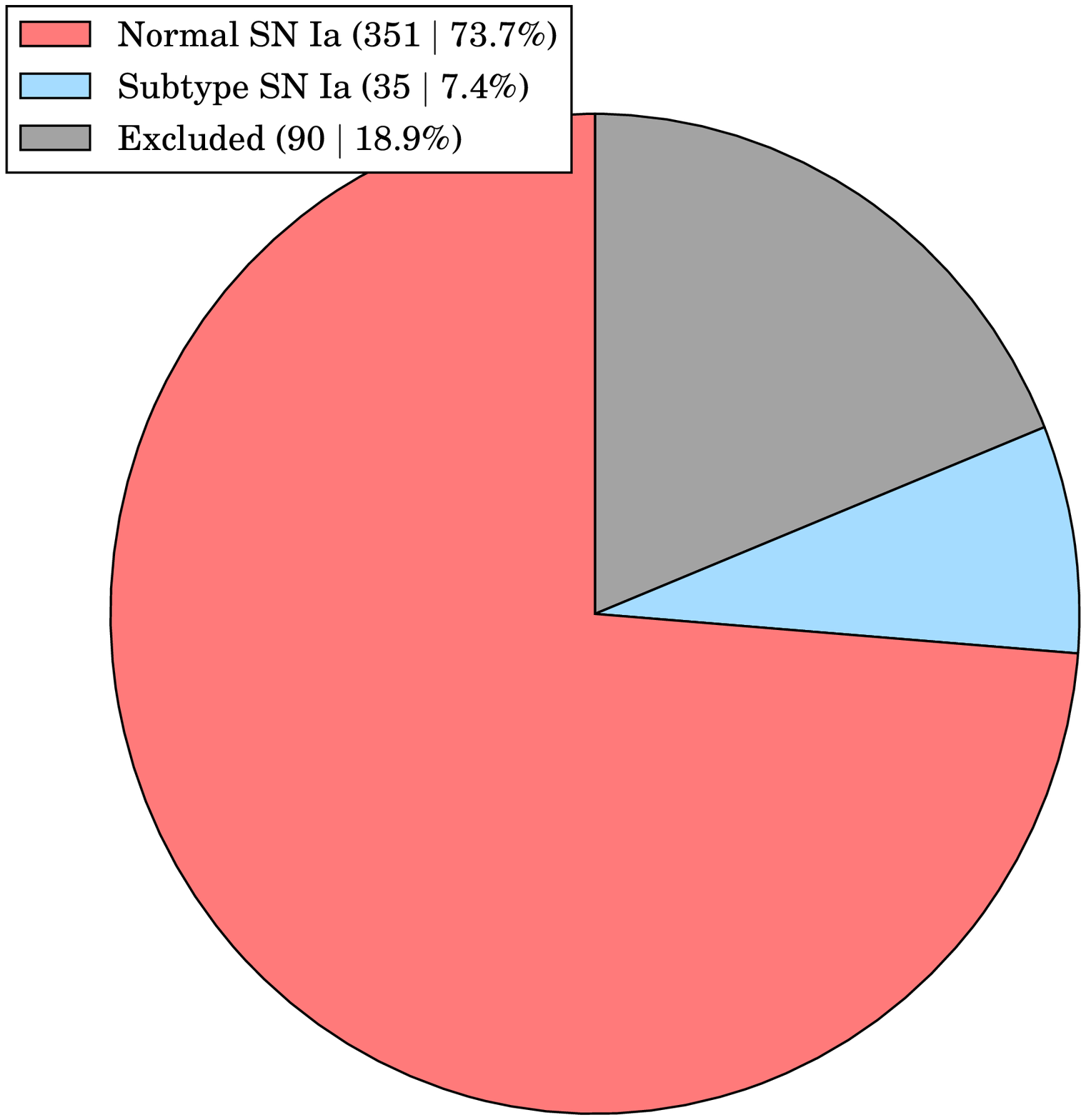}}
\centering{\includegraphics[scale=1.,width=0.49\textwidth,trim=0.pt 0.pt 0.pt 0.pt,clip]{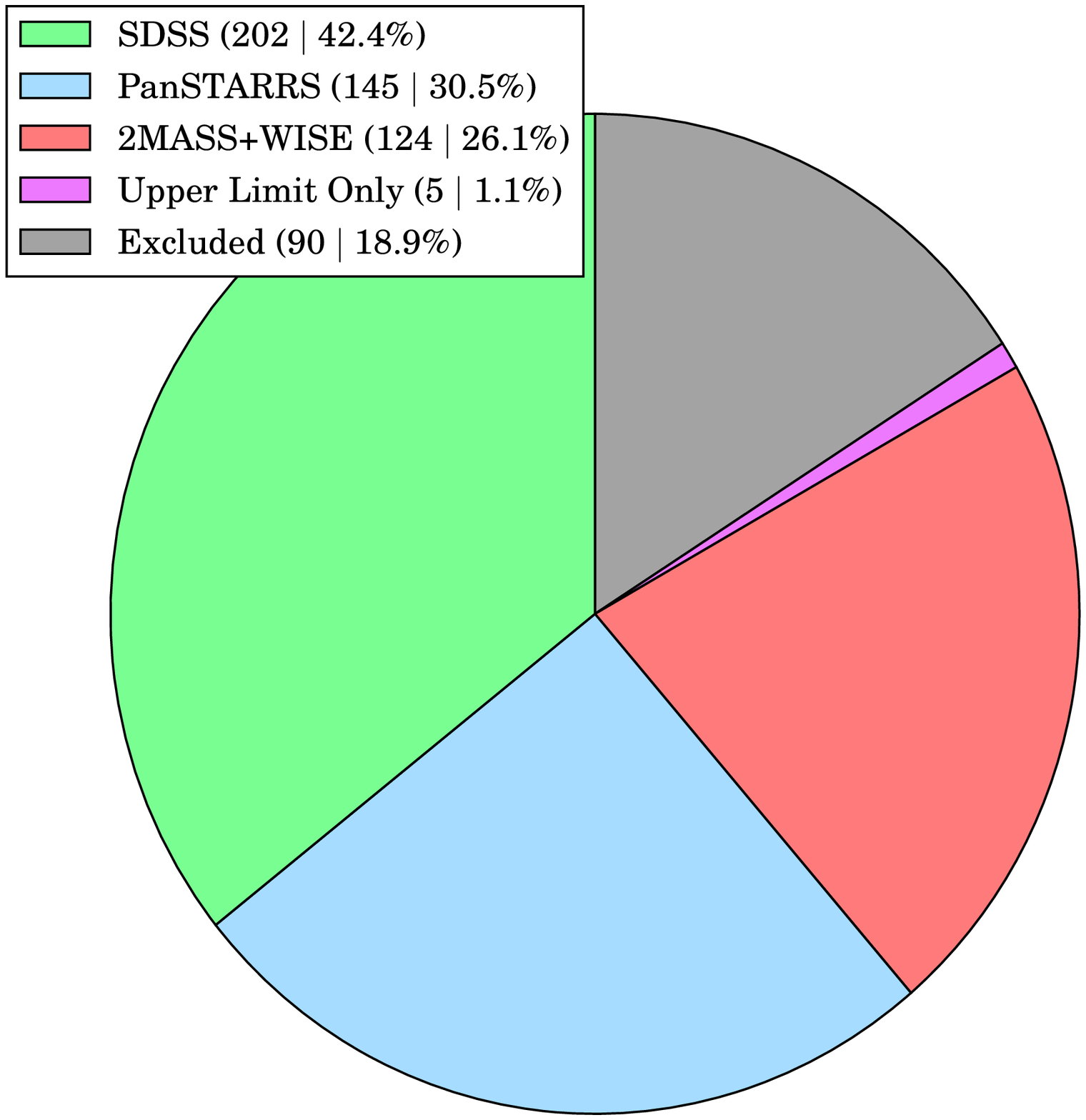}}
\caption{Left: The breakdown of the SN Ia types included in our analysis. The overall sample comes from the ASAS-SN Bright Supernova Catalogs \citep{snCat2,snCat3,snCat1}. The SNe Ia excluded from our analysis were either not recovered by ASAS-SN, or were peculiar in type. Right: The breakdown of the archival data used to model the SN Ia host galaxies. The galaxies which hosted unrecovered or peculiar SNe Ia are excluded. A small number of galaxies were lacking the data needed for a robust mass estimate; their masses should be regarded as upper limits. For every host galaxy, we incorporate the 2MASS$+$WISE data in the modeling whenever possible. When reliable SDSS photometry is not available, we use PanSTARRS stack data or, for the galaxies outside the PanSTARRS footprint ($\delta < -30^{\circ}$), the masses are estimated from the 2MASS$+$WISE data only.}
\label{fig:pie}
\end{figure*}

We assembled the archival host data from the ASAS-SN Bright Supernova Catalogs \citep[][Tables 2 and 4]{snCat2,snCat3,snCat1}. These tables contain archival data from the Galaxy Evolution Explorer \citep[\textit{GALEX};][]{Morrissey07} All Sky Imaging Survey, the optical $ugriz$ model magnitudes from the Sloan Digital Sky Survey Data Release 13 \citep[SDSS DR13;][]{Albareti17}, the near-infrared (NIR) $JHK_S$ magnitudes from the Two-Micron All Sky Survey \citep[2MASS][]{Skrutskie06}, and the IR $W1$ and $W2$ magnitudes from the \textit{Wide-field Infrared Survey Explorer} \citep[\textit{WISE;}][]{Wright10} AllWISE source catalog. In Figure~\ref{fig:pie} we show the fractional representation of SNe Ia subtypes used in this analysis (left) as well as a breakdown of the photometric survey data used to model their hosts (right).

To supplement the optical coverage of host galaxies in our sample, we also retrieve the $grizy$ data from the Panoramic Survey Telescope and Rapid Response System \citep[Pan-STARRS;][]{Chambers16, Flewelling16}, which expands the sample of SN Ia host galaxies with optical coverage to include all galaxies with declinations $\delta > -30^{\circ}$ and apparent magnitudes $m_g \lesssim 23$. In order to assemble a sample of Pan-STARRS magnitudes, we use the Pan-STARRS stack images and the Pan-STARRS Mean Object Catalog to identify the sources and their unique object IDs. We use these \textit{objIDs} to cross match with the \textit{StackObjectThin} and \textit{StackObjectAttributes} tables, and then extract the stacked Kron magnitudes and radii corresponding to the primary detection for each host galaxy. We also retrieve the detection flags, but in most cases we find that a by-eye inspection of the galaxy spectral energy distributions (SEDs) provides a more robust discriminator of the reliability of the photometry. For the host galaxies with SEDs that fail our by-eye inspection (e.g., due to vastly different Kron radii amongst the filters), or if the host was too faint or diffuse to be detected by the Pan-STARRS pipeline, we perform aperture photometry on the stacked images with a circular aperture and a fixed radius. The aperture is chosen by eye to capture as much of the galaxy flux as possible; circular apertures are also used to mask contaminating sources such as foreground stars. As a test, we compared our ``by hand'' magnitudes with reliable stacked Kron magnitudes and found good agreement.

\section{Analysis}
\label{sec:analysis}

\begin{figure*}
\centering{\includegraphics[scale=1.,width=0.49\textwidth,trim=0.pt 0.pt 0.pt 0.pt,clip]{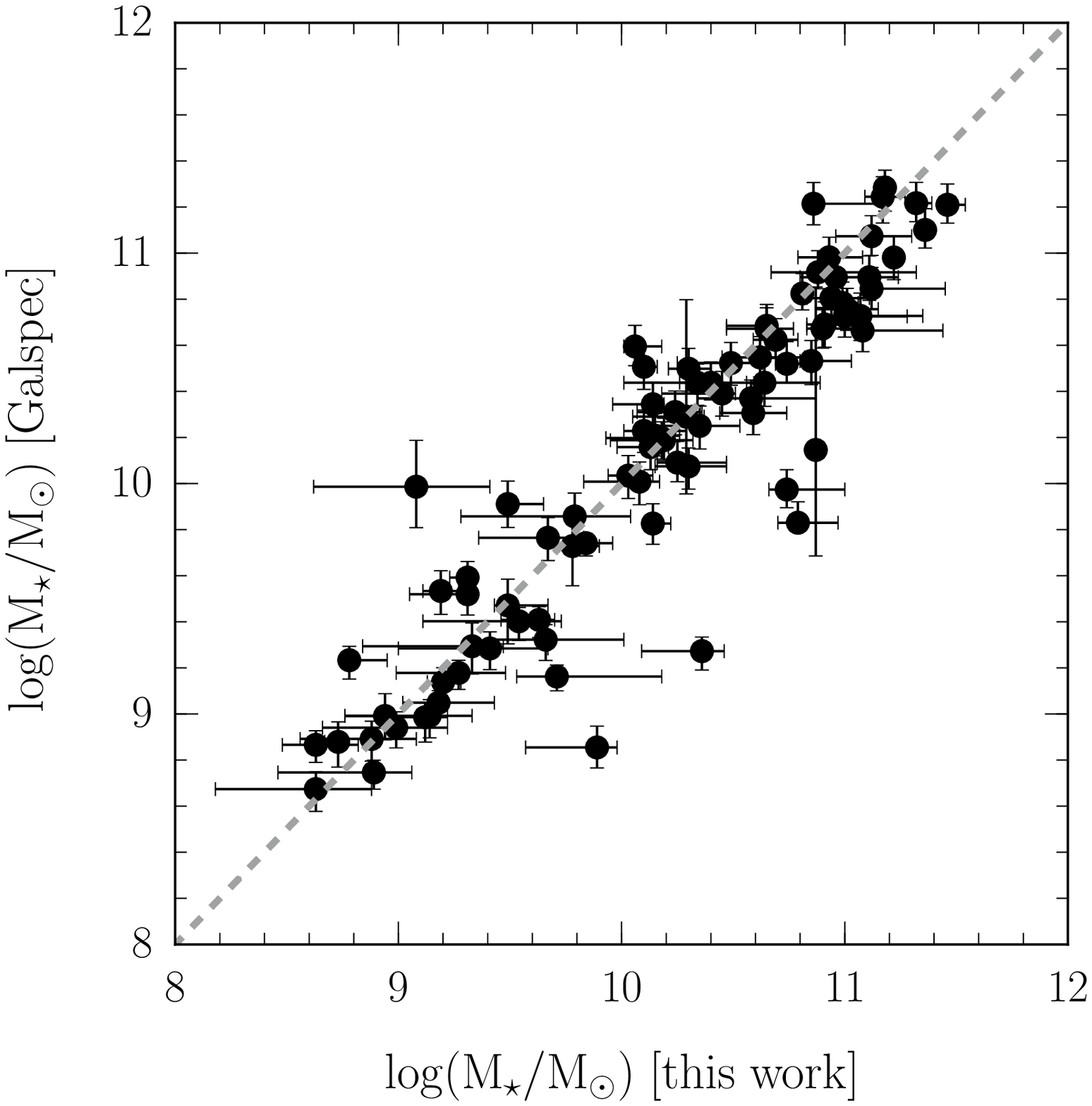}}
\centering{\includegraphics[scale=1.,width=0.49\textwidth,trim=0.pt 0.pt 0.pt 0.pt,clip]{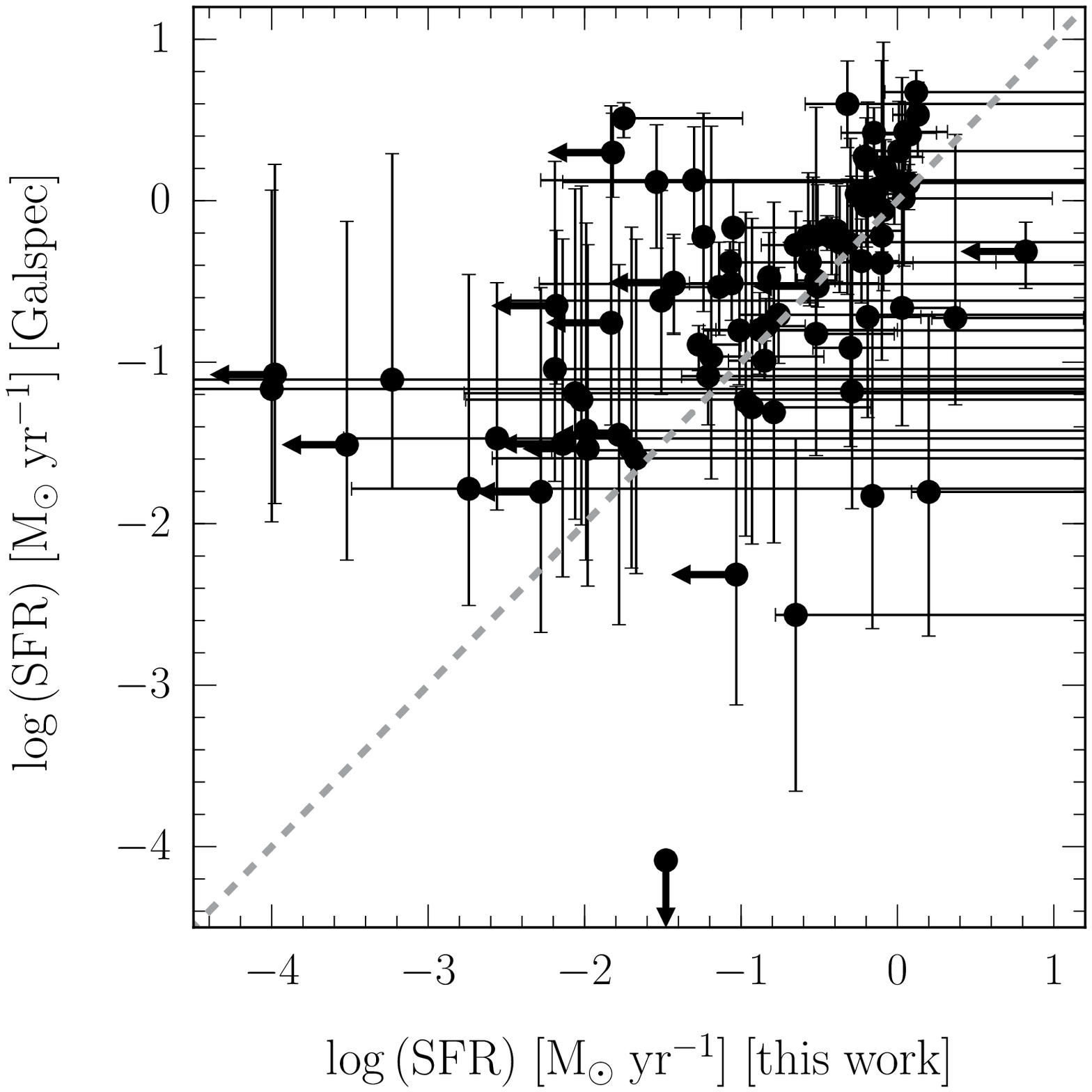}}
\caption{Stellar masses (left) and SFRs (right) derived here compared to the MPA-JHU Galspec estimates. We find generally good agreement, although our SFRs are less constrained and may be systematically lower in high SFR galaxies.}
\label{fig:galspec_comp}
\end{figure*}

\begin{figure*}
\centering{\includegraphics[scale=1.,width=0.49\textwidth,trim=0.pt 0.pt 0.pt 0.pt,clip]{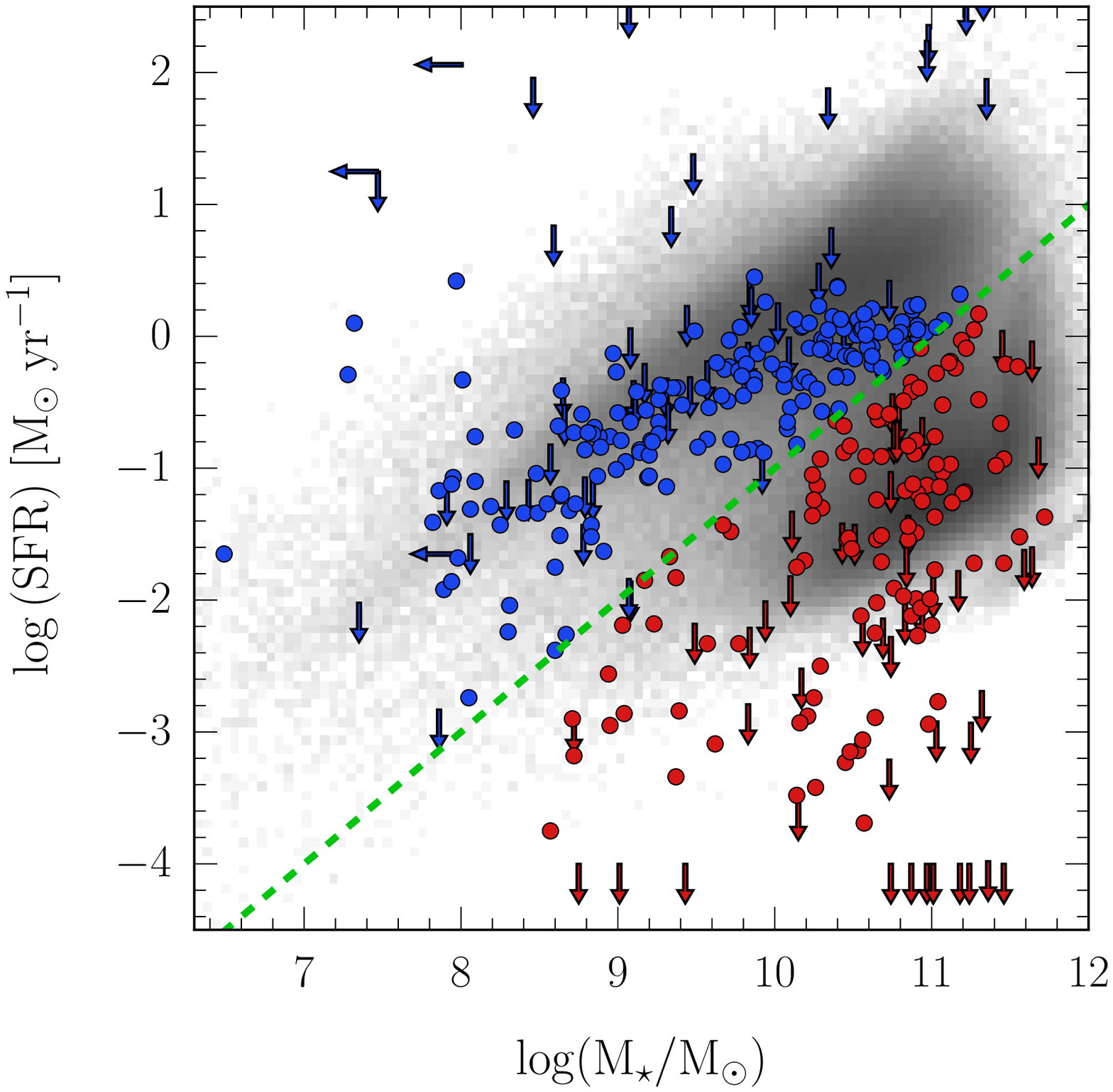}}
\centering{\includegraphics[scale=1.,width=0.49\textwidth,trim=0.pt 0.pt 0.pt 0.pt,clip]{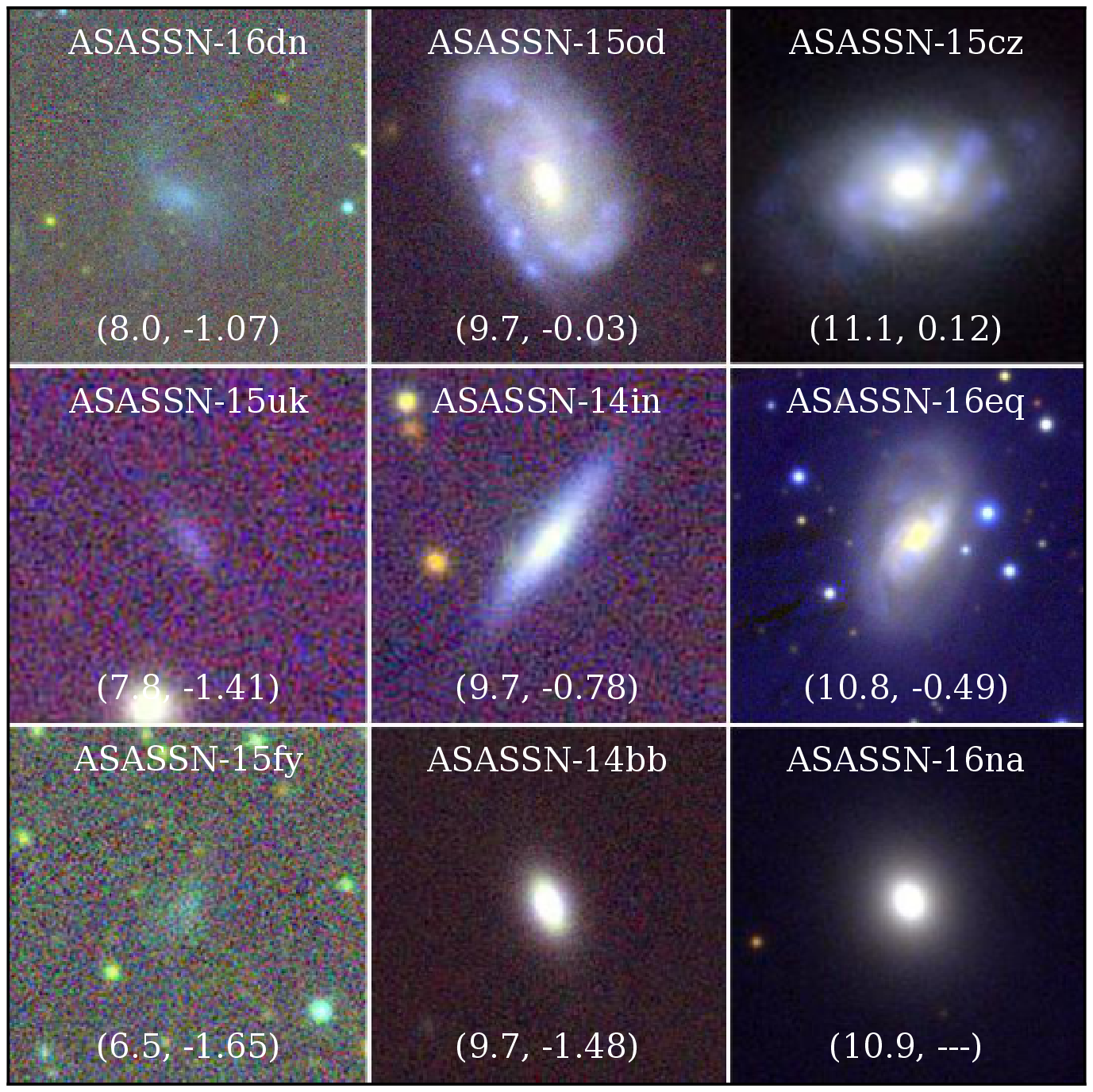}}
\caption{Left: Comparison of all host galaxies in our sample (colored symbols) with the MPA-JHU Galspec galaxies (gray contours) in the $\mstar-{\rm SFR}$ plane. The green line shows our division between actively star-forming galaxies (blue symbols) and passive galaxies (red symbols), corresponding to a ${\rm \log(sSFR)} < -11~{\rm yr}^{-1}$. Right: PanSTARRS images for a sample of host galaxies, demonstrating the diversity of our host galaxy sample. The image scale corresponds to a proper distance of $20$~kpc on a side. The numbers in parentheses are $\log(\mstar/\msun)$ and $\log({\rm SFR})$, respectively. The images are ordered such that the stellar mass is increasing rightward, and the SFR is increasing upward.}
\label{fig:galspec2D}
\end{figure*}

\begin{figure*}
\centering{\includegraphics[scale=1.,width=0.49\textwidth,trim=0.pt 0.pt 0.pt 0.pt,clip]{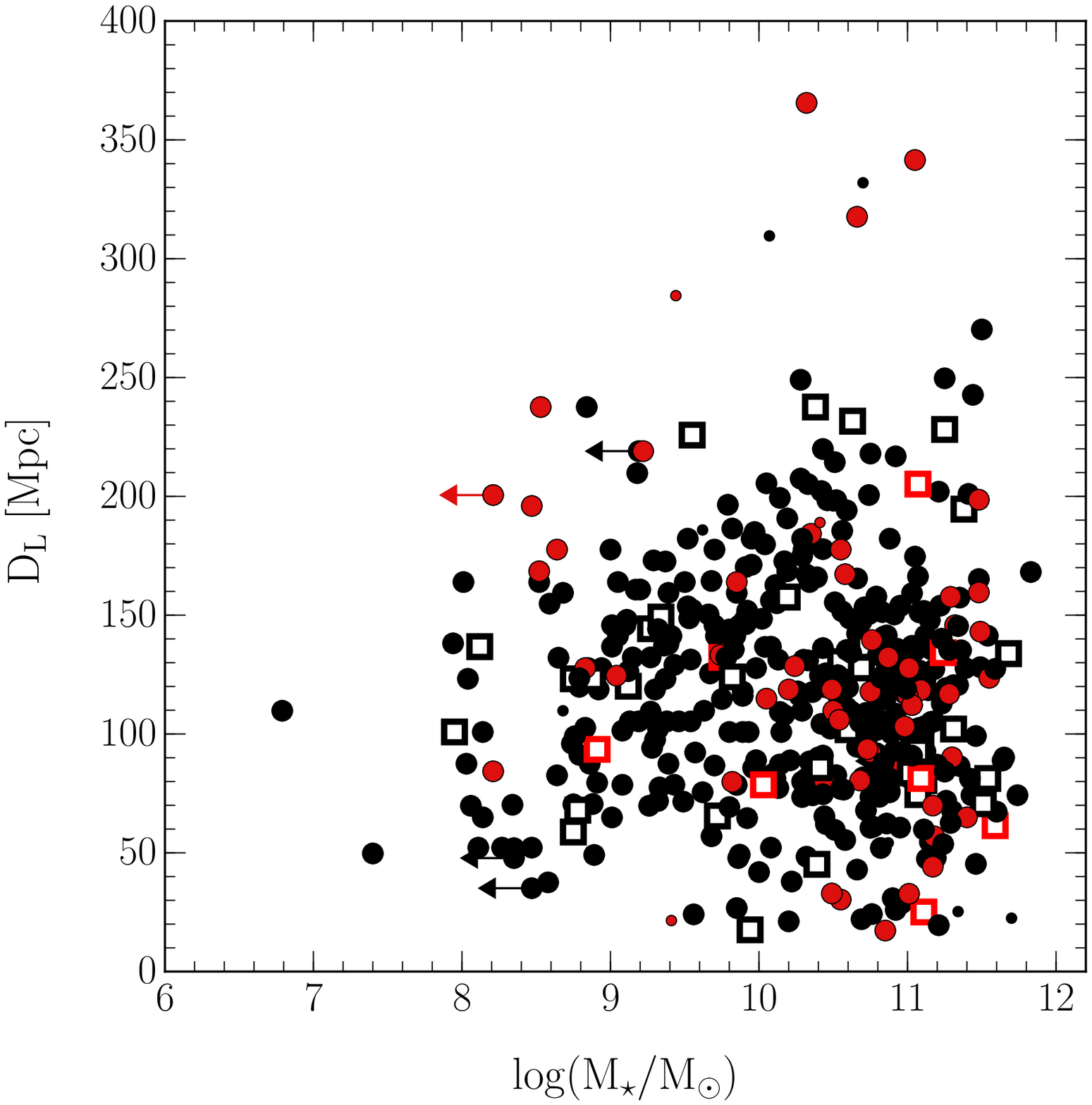}}
\centering{\includegraphics[scale=1.,width=0.49\textwidth,trim=0.pt 0.pt 0.pt 0.pt,clip]{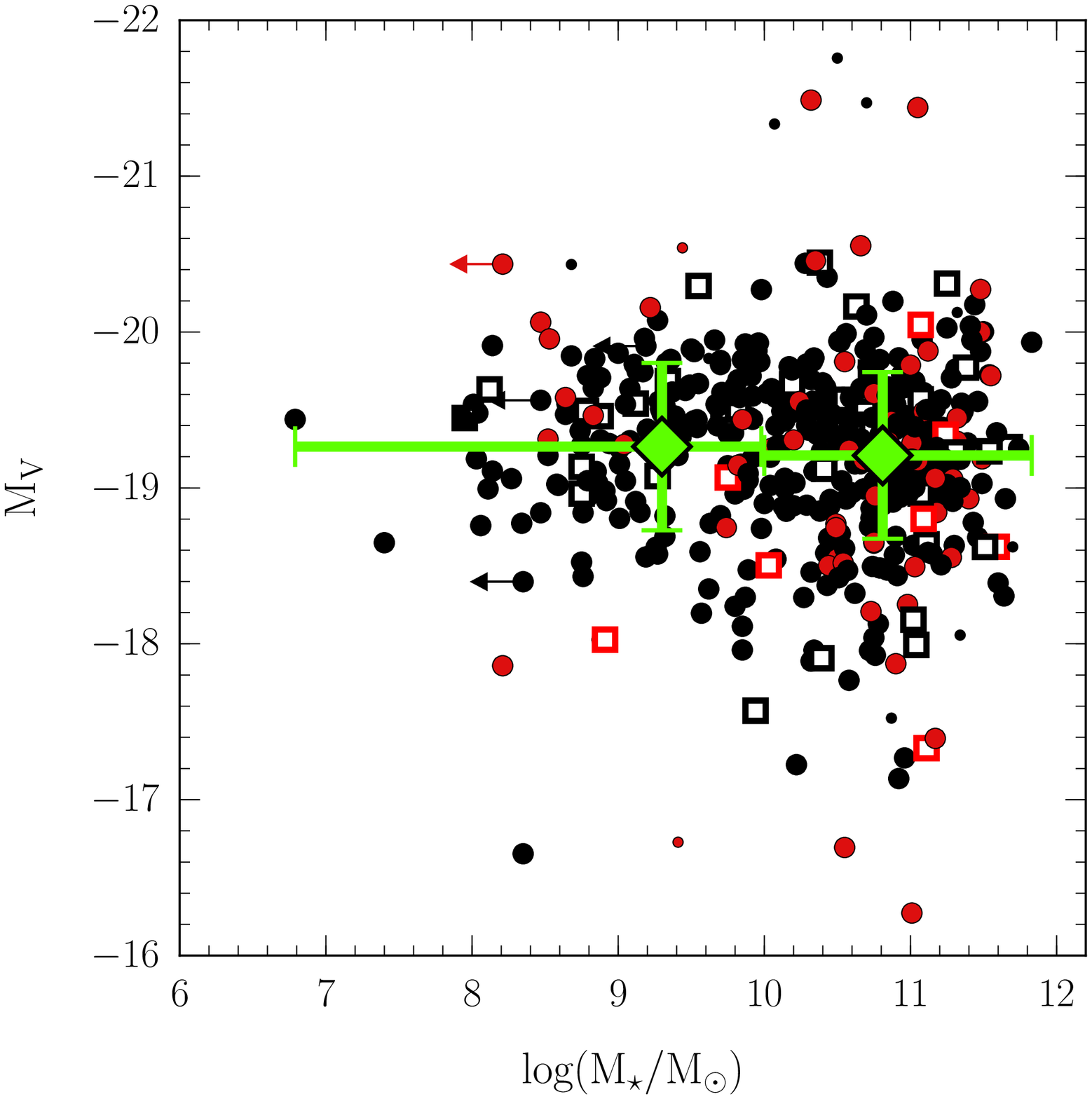}}
\caption{Left: Luminosity distance versus host galaxy stellar mass. The color and symbol scheme is the same as in Figure~\ref{fig:absMag_dist}. Right: SN Ia peak absolute magnitudes versus host galaxy stellar mass. The green diamonds show the mean absolute magnitude for galaxies hosting the discovered or recovered SNe Ia above $10^{10}\msun$ ($N=243$), and below $10^{10}\msun$ ($N=143$). The error bars show the range of masses and the standard deviation of the absolute magnitudes in the two samples. The difference in the mean absolute magnitudes of the two samples essentially consistent with zero ($0.057 \pm 0.056$ mag). These averages have not been corrected for the overrepresentation of luminous SNe inherent to a magnitude limited sample such as this one, but we note that the the same calculation performed on the smaller volume limited sample suggests no significant difference ($0.06 \pm 0.12$ mag).}
\label{fig:mass_dist_mag}
\end{figure*}

\begin{figure*}
\centering{\includegraphics[scale=1.,width=0.49\textwidth,trim=0.pt 0.pt 0.pt 0.pt,clip]{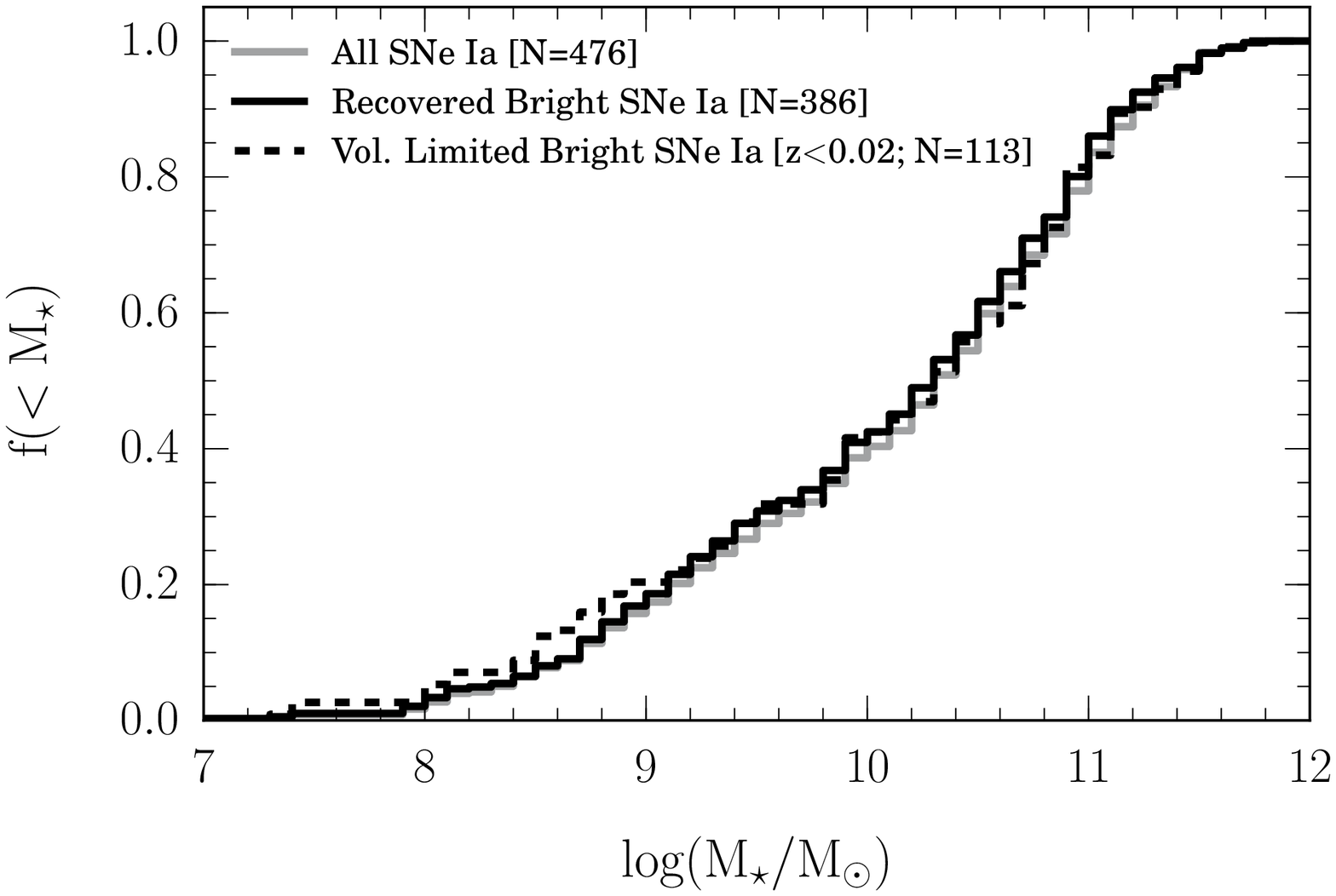}}
\centering{\includegraphics[scale=1.,width=0.49\textwidth,trim=0.pt 0.pt 0.pt 0.pt,clip]{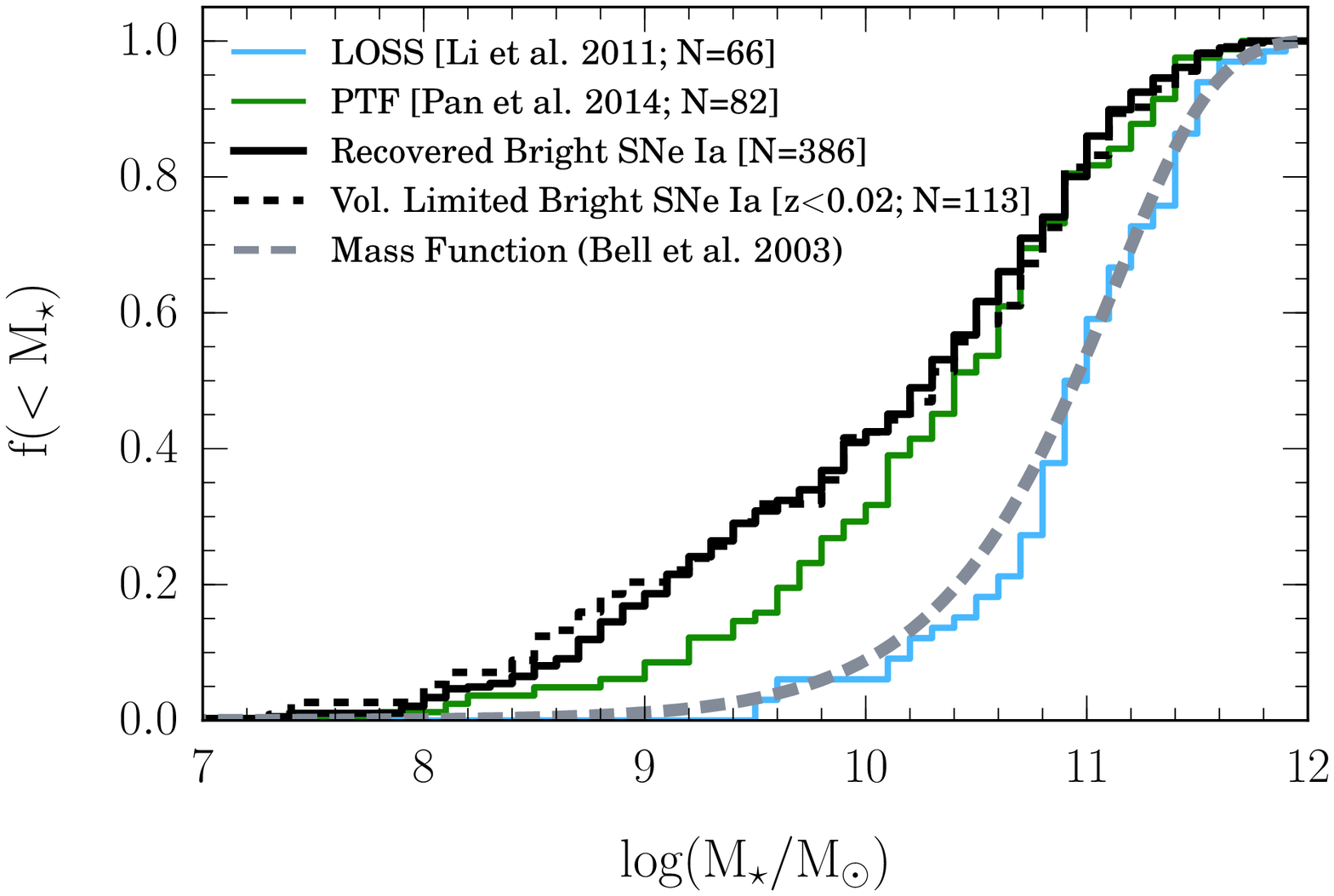}}
\caption{Left: Cumulative distribution of host galaxy masses for the entire Bright Supernova Sample (solid gray histogram), the host galaxies with SNe Ia either discovered or recovered by ASAS-SN (solid black histogram), and the low redshift host galaxies with SNe Ia either discovered or recovered by ASAS-SN (dashed black histogram). Right: Cumulative distribution of SN Ia host galaxies from various surveys. The black histograms show the host galaxies with SNe Ia discovered/recovered by ASAS-SN. The blue histogram shows the volume-limited SN Ia host galaxy sample from \citet{Loss2}, the green histogram shows the distribution of SN Ia host galaxies from PTF \citep{Pan14}, and the gray dashed line shows the cumulative galaxy mass function of all galaxies from \citet{Bell03}. The ASAS-SN sample includes a larger fraction of low-mass galaxies than other low redshift SN surveys.}
\label{fig:mass_dist_basic}
\end{figure*}

\begin{figure*}
\centering{\includegraphics[scale=1.,width=\textwidth,trim=0.pt 0.pt 0.pt 0.pt,clip]{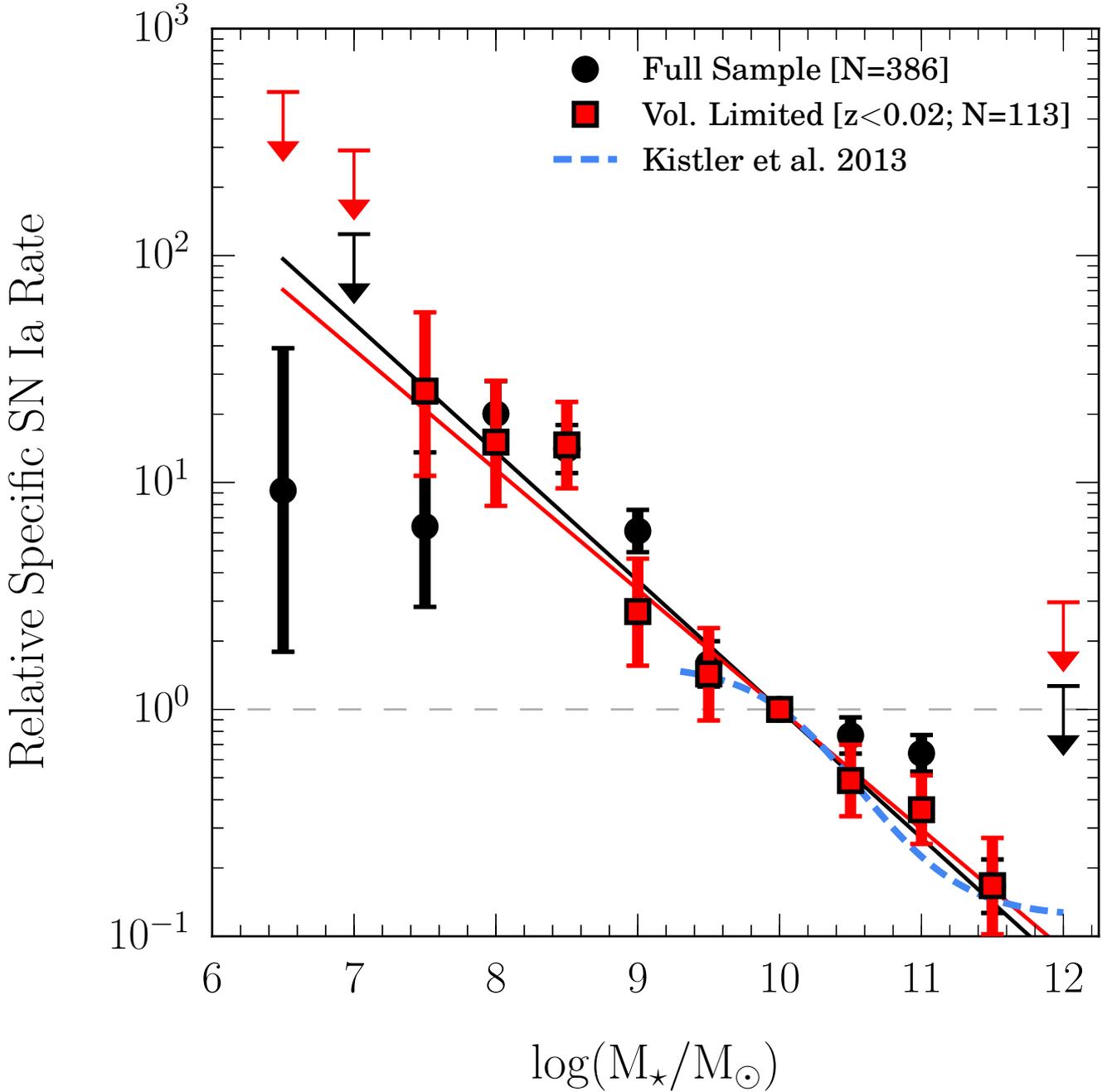}}
\caption{The SN Ia rate per unit stellar mass as a function of host galaxy mass, relative to the rate at $10^{10}\msun$. The black points are calculated using the full sample of host galaxies with SNe Ia either discovered or recovered by ASAS-SN; the red points are calculated from the volume-limited sample. Error bars correspond to the $84\%$ confidence intervals computed from the \citet{Gehrels86} approximations for binomial statistics. The dashed blue curve shows the \citet{Kistler13} analytic fit to the LOSS SN Ia host galaxy sample in \citet{Loss3}. The black and red solid lines show the approximate dependence of the specific \snia\ rate, assuming a power law normalized to unity at $10^{10}\msun$.}
\label{fig:rel_rate}
\end{figure*}

\begin{figure}
\centering{\includegraphics[scale=1.,width=0.49\textwidth,trim=0.pt 0.pt 0.pt 0.pt,clip]{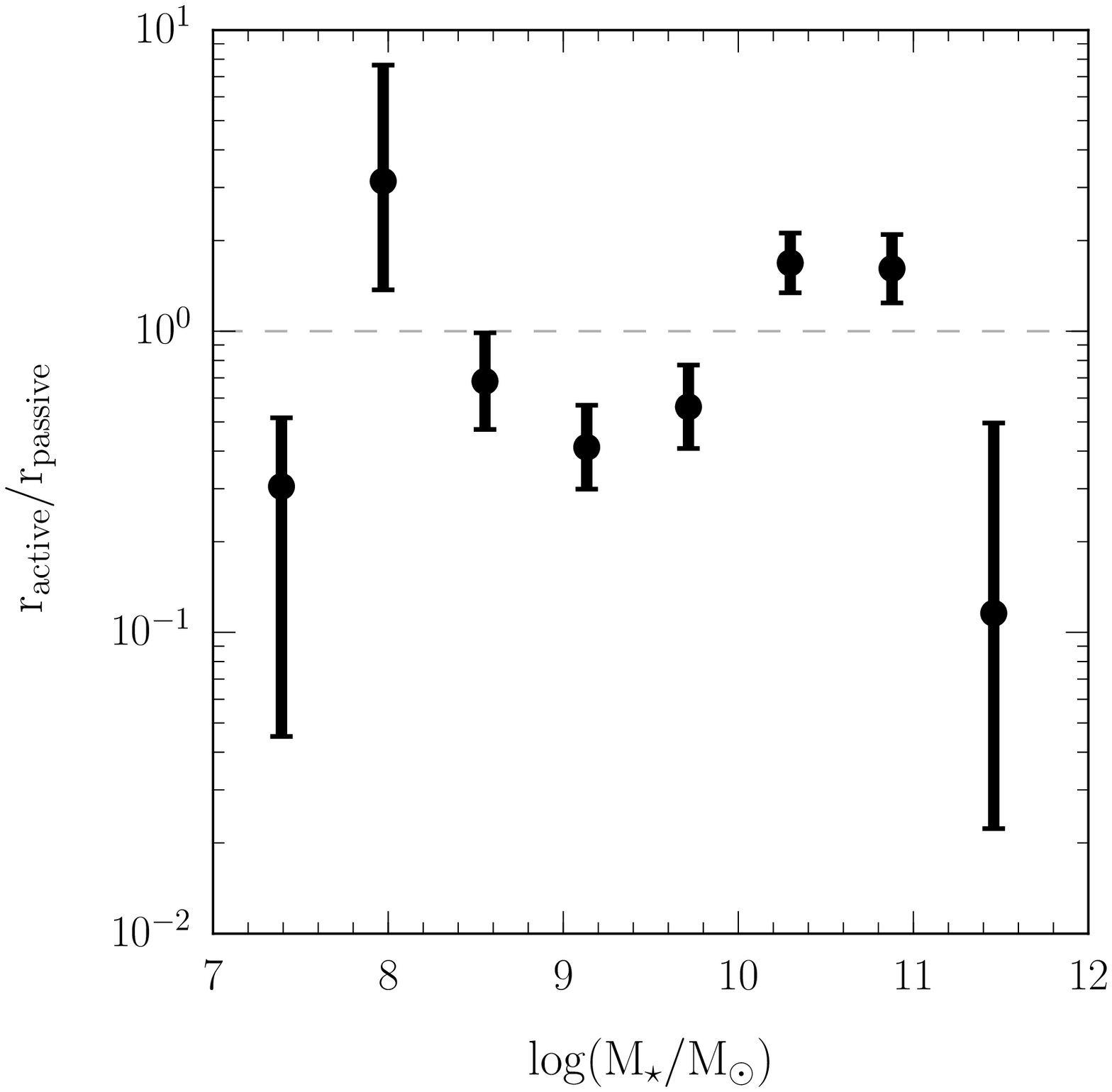}}
\caption{The ratio of the \snia\ rate in actively star-forming galaxies to that in passive galaxies, as a function of host galaxy stellar mass.}
\label{fig:rel_rate_sfr}
\end{figure}

The 2MASS and WISE magnitudes form the starting point for the host galaxy SED modeling. The WISE catalog, in particular, is relatively deep, and nearly all host galaxies are detected in both W1 and W2. These bands are primarily sensitive to the stellar mass. Similarly, we also include the GALEX NUV magnitudes in the model by default, since the NUV coverage provides constraints on the stellar age and mass-to-light ratio. However, there is no uniform all-sky optical survey, so we have a choice as to which dataset to include in the SED fitting. When possible, we adopt the masses derived from the SDSS data. In a small number of instances, the SDSS magnitudes are unreliable and we adopt the mass derived from the Pan-STARRS data instead. In a few cases, the host galaxy spans several arcminutes on the sky, and the catalogued photometry for most surveys does not reflect the true photometry of the host galaxy. In these instances, we adopt the $JHK_S$ magnitudes from the 2MASS Large Galaxy Atlas \citep[2MASS LGA;][]{Jarrett03} and model the galaxy using only these magnitudes.

We model the host galaxy SEDs with the publicly available Fitting and Assessment of Synthetic Templates \citep[\textsc{fast}][]{Kriek09}. We assume a \citet{Cardelli89} extinction law with $R_V = 3.1$ and the Galactic extinction taken from \citet{Schlafly11}. We also assume an exponentially declining star-formation history, a Salpeter initial mass function, and the \citet{Bruzual03} stellar population models. Given the heterogenous mix of photometry for these galaxies, the statistical uncertainties are generally much smaller than the systematic uncertainties. The statistical uncertainties of a given measurement may be a few hundredths of a magnitude or smaller, while the systematic uncertainties (e.g., due to variations in aperture size or background subtraction between filters) are generally at the tens of percent level or more. Artificially small uncertainties in a single survey or measurement will drive the fit to match that particular survey at the expense of fitting the archival data from other archival surveys. In order to minimize this effect, we assign a minimum uncertainty of $0.1$ mag to the photometry when modeling the host galaxy SED.

For each galaxy we inspect the various fits to the archival data and adopt a preferred mass given the considerations outlined above. As a check, we compare the masses derived with our \textsc{fast} fits to those from the MPA-JHU Galspec pipeline \citep{Kauffmann03} where available, and we find the results are generally consistent. As shown in Figure~\ref{fig:galspec_comp}, our SFR estimates are less well constrained than the Galspec data, but this is to be expected given that we are utilizing photometric data only. Additionally, while our SFR estimates agree reasonably well with the Galspec values, there is some indication that for high SFR galaxies, our SFRs are systematically lower than the Galspec values. 

We also see an indication of discrepant SFR estimates when we view the distribution of SDSS galaxies compared to the distribution of our host galaxies in the $\mstar-{\rm SFR}$ plane. This is shown in Figure~\ref{fig:galspec2D}. The gray shaded contours represent the SDSS galaxies, while the colored circles show host galaxies from our sample. The green dashed line corresponding to a ${\rm\log(sSFR)} = -11~{\rm yr}^{-1}$ represents our nominal separation between high and low star-forming galaxies. For host galaxies with nominal SFR $<10^{-4}~\msun~{\rm yr}^{-1}$, we adopt the SFR upper limit from the \textsc{fast} results, which are shown with downward arrows. The host galaxies from our sample follow the overall distribution of SDSS galaxies, but at the high SFR end, we appear to have a deficit of SN Ia hosts. However, given the generally weak constraints on our SFR estimates and the potential for significant systematics between our models and the Galspec results, we refrain from making any strong claims about the origin of this trend. Future work addressing the SFR as well as chemical abundances of SN Ia host galaxies will require additional observational constraints (namely spectroscopic follow-up) in order to 1) improve the constraints on these parameters, and 2) reduce the potential for modeling biases between our host galaxy population and the galaxy population at large.

In Figure~\ref{fig:mass_dist_mag} we show the distribution of host galaxy masses versus luminosity distance (left) and SN absolute magnitude (right). The color and symbol schemes are the same as Figure~\ref{fig:absMag_dist}. Leftward arrows denote upper limits on the host galaxy mass. The green diamonds in the right panel of Figure~\ref{fig:mass_dist_mag} show the mean host galaxy mass and SN Ia absolute magnitude for galaxies above $10^{10}\msun$ ($N=243$) and below $10^{10}\msun$ ($N=143$) for the full sample used in this analysis. The difference in the median absolute magnitudes of the two samples is $\sim 0.14$~mag, while the difference in the mean absolute magnitudes is smaller ($0.057 \pm 0.056$ mag), suggesting the SN Ia in more massive galaxies are marginally fainter than the SNe Ia in the lower mass galaxies. This is consistent with previous studies \citep[e.g.,][]{Sullivan10,Pan14}, which found that low stretch (i.e., fainter) SNe Ia are found in more massive galaxies. These averages have not been corrected for the overrepresentation of luminous SNe inherent to a magnitude limited sample such as this one, but we note that the the same calculation performed on the smaller volume limited sample suggests no significant difference ($0.06 \pm 0.12$ mag). The overall amplitude of this effect is not large, and would only result in a $\lesssim15\%$ enhancement of low-mass galaxies over high-mass galaxies, all else being equal.

In Figure~\ref{fig:mass_dist_basic} we show the cumulative distribution of host galaxy masses. In the left panel, we show all 476 galaxies in the ASAS-SN Bright Supernova Catalogs (gray solid line) the 386 galaxies hosting SNe Ia either discovered or recovered by ASAS-SN (black solid line), and the volume-limited ($z\leq0.02$) subset of 113 galaxies hosting SNe Ia either discovered or recovered by ASAS-SN. The mass distributions are quite similar, but the volume-limited sample has a slight enhancement of low-mass galaxies compared to the full sample of discovered/recovered SN Ia hosts. 

In the right panel we compare the total and volume-limited samples from ASAS-SN (black solid and dashed histograms, respectively) to the distribution of host galaxy masses from other low redshift supernova surveys. The green and blue histograms show the distribution of SN Ia host galaxy masses from PTF \citep{Pan14} and the volume-limited LOSS sample \citep{Loss2}, respectively. The gray dashed line shows the overall galaxy mass function from \citet{Bell03}. The host galaxy mass distributions from the ASAS-SN sample (and to a lesser extent the PTF sample) are much more sensitive to lower mass galaxies than LOSS. This is principally due to the fact that LOSS was a targeted SN survey and preferentially observed large galaxies. The LOSS survey was instrumental in demonstrating that lower mass galaxies produce a larger number of SNe Ia than higher mass galaxies, but LOSS was only able to measure the \snia\ rate in galaxies with stellar masses $\gtrsim10^{9}\msun$ \citep{Loss3}. It is important to note that this effect is not due a bias in ASAS-SN against SNe in massive galaxies. The results in \citet{snCat3} demonstrate that ASAS-SN is actually more sensitive to SNe in large galaxies (including SNe near the galaxy nuclei) than other surveys. Additionally, the ASAS-SN sample has a significantly larger number of SNe than the other surveys. The benefit of using an unbiased SN survey to conduct a census of SN host galaxies is quite clear; SNe occur in locations that, in practice, cannot be monitored with traditional targeted surveys. 

In order to obtain the observed specific SN Ia rate, we need to apply several corrections to the host galaxy mass distribution, and we follow the same procedure used to compute the relative luminosity function. First, we need to increase the weights of the galaxies hosting faint SNe. We compute these weights from the completeness corrections presented in Section~\ref{sec:data}. Additionally, we need to correct for time dilation and the fact that the luminous SNe can be observed to greater distances. We find that there is at most only a modest correlation between the SN luminosity and host galaxy stellar mass, and thus do not implement any explicit correction for this relationship. In short, we use Equation~\ref{eqn:sum} to sum over the host galaxies in each stellar mass bin, using the properties of the individual SNe to compute the appropriate weights. Finally, we need to assume a form for the underlying galaxy stellar mass function. We adopt the $g$-band derived stellar mass function from \citet{Bell03}, which we have converted to a Salpeter IMF by scaling their masses by $-0.15$ dex. For each bin, we divide the weighted histogram by the integral of the stellar mass function over the width of the bin (i.e., $\int_{\rm M_{1}}^{M_{2}} M dM dn/dM$) and normalize to the $10^{10}\msun$ bin to obtain the relative SN Ia rate per unit stellar mass. We note that an enhanced specific SN Ia rate at low masses could potentially be due to an underestimation of the stellar mass function at low masses. However, the results from \citet{Baldry12} show that the faint-end slope of the stellar mass function remains relatively constant even down to the lowest masses considered here.

Figure~\ref{fig:rel_rate} shows this normalized specific SN Ia rate as a function of host galaxy mass over the range $6.25 \leq \log(M_{\star}/M_{\odot}) \leq 12.25$. The black circles show the rate calculated using the full ASAS-SN sample, and the red squares show the rate calculated from the volume-limited sample. The dashed blue line shows the analytic fit to the \citet{Loss3} results from \citet{Kistler13}. At the high mass end ($9.5 \lesssim \log(M_{\star}/M_{\odot}) \lesssim 11.5$), we find remarkably good agreement with the results from \citet{Kistler13}. Moving towards lower masses, the \citet{Kistler13} curve suggests a flattening in specific SN Ia rate, but this is largely due to their assumed analytic form. The ASAS-SN data show that the specific SN Ia rate continues to rise towards lower mass galaxies, down to stellar masses of $\sim10^7\msun$. That is, progressively lower mass galaxies produce more SNe Ia per unit mass than more massive galaxies. This trend is broadly consistent with previous studies \citep{Sullivan06,Loss3,Smith12,Graur13}, but this has never been shown for low redshift galaxies spanning such a broad range of masses (e.g., LOSS is only sensitive to galaxies with stellar masses $\mstar \gtrsim 10^{9}\msun$). For both the full and volume-limited samples, we fit a power law to the bins above $10^8\msun$ normalized to unity at $10^{10}\msun$, such that $r = (M/10^{10}\msun)^{\alpha}$. We find $\alpha_{\rm full} = -0.57 \pm 0.09$ and $\alpha_{\rm vol.} = -0.53 \pm 0.05$, which are shown by the black and red lines for the full sample and volume-limited subsample, respectively. These values are in excellent agreement with the results from \citet{Loss3}.

Compared to the higher redshift surveys, we find a steeper mass dependence. \citet{Sullivan06} find that the specific rate is essentially flat for the passive galaxy population, and a similar trend is seen for the SDSS sample in \citet{Smith12}. Instead, these studies have argued that the increased rate of SN Ia at low stellar masses is primarily due to increasing sSFR (or decreasing age) with decreasing stellar mass \citep[e.g.,][]{Mannucci05,Sullivan06,Smith12}. While there is clear evidence that sSFR increases with decreasing stellar mass \citep{Brinchmann04,Speagle14}, the relationship in the low redshift galaxy population may be more shallow than previously believed \citep{Boogaard18}, and thus would not explain the enhanced SN Ia rates at low masses we observe here. It is quite likely that host galaxy metallicity, which is also highly correlated with stellar mass \citep[e.g.,][]{Tremonti04,Gallazzi05}, is an important factor driving SN demographics \citep{Prieto08,Khan11a,Khan11b,Kistler13}. It is critical that we understand the SFRs and metallicities of the host galaxy population if we are to understand the physical origin of the enhanced \snia\ rate at low stellar masses.

Given the relatively loose constraints on the host galaxy SFRs, we simply split our sample into a star-forming subsample (${\rm \log(sSFR)}>-11~{\rm yr}^{-1}$) and a passively evolving sample (${\rm \log(sSFR)}\le-11~{\rm yr}^{-1}$). Galaxies with only upper limits on SFRs are automatically assigned to the passive sample. We compute the relative rates following the same procedure as outlined above, but include an additional factor in the weights accounting for the fractions of blue and red galaxies as a function of stellar mass. At low masses ($\mstar \lesssim 10^9 \msun$) the galaxy population is dominated by star-forming galaxies with relatively few passive galaxies, while at high masses the opposite is true. We derive the correction factors by computing the fraction of early and late type galaxies the stellar mass functions presented in \citet{Bell03}. Our procedure assumes that the mapping in \citet{Bell03} between early and late type galaxies is reasonably consistent with our division between passive and star-forming galaxies. We note, however, that \citet{Baldry12} argue that below $\sim 10^{8.5}\msun$, the relative populations of blue and red galaxies are not all that well constrained, and that at very low masses (e.g., $\lesssim10^{7.5}\msun$) the fractions of red and blue galaxies may be comparable. Given these uncertainties, we adopt a maximum correction factor for the low-mass galaxies fixed to the value computed at $10^{9}\msun$. 

We show the \snia\ rate in actively star-forming galaxies relative to that in passive galaxies in Figure~\ref{fig:rel_rate_sfr}. We find no strong evidence that the relative \snia\ rate strongly depends on star formation activity. In the high-mass galaxies where most of the \sneia\ originate ($\mstar \sim 10^{10} - 10^{11}\msun$), the actively star-forming galaxies appear to produce marginally more \sneia\ than the passive galaxies of the same mass. This is consistent with previous findings \citep[e.g.,][]{Mannucci05,Graur17}. On the other hand, at lower masses, there is no evidence that active galaxies are more efficient at producing \sneia\ than passive galaxies. In fact, in both cases there is little evidence that the ratio of the relative rates in the two samples differs from unity. In any case, a careful assessment of this behavior requires more robust measurements of SFRs, and better estimates of the relative galaxy stellar mass functions.

\section{Conclusions}
\label{sec:conclusions}

We leveraged the statistical power of three years of discoveries presented in the ASAS-SN Bright Supernova Catalogs to construct a sample of SNe Ia that is largely unbiased with respect to host galaxy properties and nearly complete within $\sim100$~Mpc. We derived the relative completeness of the $V$-band component of ASAS-SN as a function of apparent magnitude which, combined with the the peak $V$-band magnitudes from the catalogs, we used to construct the relative luminosity function of \sneia\ in the low redshift universe. This luminosity function is reasonably well described by a \citet{Schechter76} function with a faint-end slope $\alpha \simeq 1.5$ and a ``knee'' at $\mstar \simeq -18.0$. 

We used archival photometric data from the near-UV, optical and near-IR wavelengths to derive masses and star formation rates for the \snia\ host galaxy population. Finally, we used this host data in conjunction with the individual SN data to derive the relative \snia\ rate as a function of host galaxy properties.

We find that the observed specific \snia\ rate scales approximately as $(\mstar/10^{10}\msun)^{-0.5}$ over 5 decades in mass from $\mstar \simeq 10^7 \msun$ to $\mstar \simeq 10^{12} \msun$. The lowest mass galaxies produce $\sim100$ times more \sneia\ per unit stellar mass than their massive counterparts. We find no strong evidence of a dependence of the specific \snia\ rate on star formation activity, but to derive meaningful constraints on the the host galaxy sSFRs, spectroscopic observations are needed. Such observations would also provide the data needed to measure chemical abundances and characterize how the \snia\ rate depends on metallicity. Given the low redshift nature of the sample, obtaining optical photometry and medium resolution spectroscopy for $\sim100\%$ of the host galaxies would be a large but feasible undertaking, especially for a volume-limited or other well-defined subsamples. Similarly, multiple efforts are underway to monitor low redshift SN Ia (\citealt{Foley18}, Chen et al., in preparation), which will be invaluable for studying the relationships between SN Ia light curve properties and their host galaxies in this revolutionary SN Ia sample. 

\section*{Acknowledgements}

The authors thank the referee for a constructive report that improved the quality of this work.

JSB, KZS, and CSK are supported by NSF grants AST-1515876 and AST-1515927. 

ASAS-SN is funded in part by the Gordon and Betty Moore Foundation through grant GBMF5490 to the Ohio State University, NSF grant AST-1515927, the Mt. Cuba Astronomical Foundation, the Center for Cosmology and AstroParticle Physics (CCAPP) at OSU, the Chinese Academy of Sciences South America Center for Astronomy (CASSACA), and the Villum Fonden (Denmark). Development of ASAS-SN has been supported by NSF grant AST-0908816, the Center for Cosmology and AstroParticle Physics at the Ohio State University, the Mt. Cuba Astronomical Foundation, and by George Skestos.

Support for JLP is provided in part by the Ministry of Economy, Development, and Tourism’s Millennium Science Initiative through grant IC120009, awarded to The Millennium Institute of Astrophysics, MAS.

JFB is supported by NSF grant No. PHY-1714479.

MDS is supported by a research grant 13261 (PI Stritzinger) from the Villum FONDEN. 

Observations made with the NASA Galaxy Evolution Explorer (GALEX) were used in the analyses presented in this manuscript. Some of the data presented in this paper were obtained from the Mikulski Archive for Space Telescopes (MAST). STScI is operated by the Association of Universities for Research in Astronomy, Inc., under NASA contract NAS5-26555. Support for MAST for non-HST data is provided by the NASA Office of Space Science via grant NNX13AC07G and by other grants and contracts.

Funding for SDSS-III has been provided by the Alfred P. Sloan Foundation, the Participating Institutions, the National Science Foundation, and the U.S. Department of Energy Office of Science. The SDSS-III web site is http://www.sdss3.org/.

This publication makes use of data products from the Two Micron All Sky Survey, which is a joint project of the University of Massachusetts and the Infrared Processing and Analysis Center/California Institute of Technology, funded by NASA and the National Science Foundation.

This publication makes use of data products from the Wide-field Infrared Survey Explorer, which is a joint project of the University of California, Los Angeles, and the Jet Propulsion Laboratory/California Institute of Technology, funded by NASA.

This research was made possible through the use of the AAVSO Photometric All-Sky Survey (APASS), funded by the Robert Martin Ayers Sciences Fund.

This research has made use of the NASA/IPAC Extragalactic Database (NED), which is operated by the Jet Propulsion Laboratory, California Institute of Technology, under contract with the National Aeronautics and Space Administration.

\bibliography{ms}
\bsp	

\appendix
\section{Host Data}
The auxilliary data for the SNe and host galaxies used in this analysis are presented in Table~\ref{tab:host_table} and the photometry of the host galaxies is presented in Table~\ref{tab:phot_table}.

\begin{landscape}
\input{./tab1_sample.tex}
\end{landscape}

\begin{landscape}
\input{./tab2_sample.tex}
\end{landscape}


\end{document}

%% file: tab1_sample.tex
\begin{table}
\begin{minipage}{\textwidth}
\caption{Physical host data.\hfill}\begin{tabular}{llcclllrr}
\hline
SN name & Host name & Included?$^{a}$ & SN type & Redshift &$V_{\rm peak}^{b}$ & $A_{\rm V}$ & $\log({\rm M}_{\star})$ [M$_{\odot}$] & $\log({\rm SFR})$ [M$_{\odot}$~yr$^{-1}$]$^{c}$  \\
\hline
ASASSN-13an & 2MASX J13453653-0719350 & True & Ia & 0.0216 & 15.8 & 0.11 & $10.52_{-0.29}^{+0.14}$ & $ 0.01_{-0.09}^{+0.07}$ \\
ASASSN-13ar & VV 478 & True & Ia & 0.01775 & 14.8 & 0.15 & $9.26_{-0.14}^{+0.07}$ & $-0.65_{-3.35}^{+0.18}$ \\
ASASSN-13av & NGC 7068 & True & Ia & 0.01729 & 15.7 & 0.27 & $10.79_{-0.18}^{+0.09}$ & $<-0.51$ \\
ASASSN-13aw & CGCG 252-043 & True & Ia & 0.016835 & 15.0 & 0.07 & $9.84_{-0.15}^{+0.45}$ & $-0.36_{-0.23}^{+0.25}$ \\
ASASSN-13bb & UGC 01395 & True & Ia & 0.017405 & 15.7 & 0.20 & $10.80_{-0.24}^{+0.08}$ & $-0.06_{-0.23}^{+0.03}$ \\
ASASSN-13cc & NGC 1954 & True & Ia & 0.01044 & 15.0 & 0.39 & $11.46_{-0.01}^{+0.01}$ & $<-4.00$ \\
ASASSN-13ch & CGCG 023-030 & True & Ia & 0.01646 & 15.8 & 0.33 & $9.12_{-0.10}^{+0.20}$ & $-0.42_{-0.28}^{+0.11}$ \\
ASASSN-13cj & CGCG 051-075 & True & Ia & 0.018 & 15.3 & 0.20 & $9.43_{-0.01}^{+0.01}$ & $<-4.00$ \\
ASASSN-13cp & ARK 477 & True & Ia & 0.023576 & 15.9 & 0.16 & $11.12_{-0.33}^{+0.01}$ & $-0.19_{-0.34}^{+0.01}$ \\
ASASSN-13cu & VIII Zw 035 & True & Ia & 0.0272 & 16.6 & 0.09 & $10.30_{-0.17}^{+0.15}$ & $-0.02_{-0.15}^{+0.02}$ \\
ASASSN-13dd & NGC 2765 & True & Ia & 0.01255 & 15.2 & 0.09 & $11.01_{-0.14}^{+0.14}$ & $<-1.83$ \\
ASASSN-13dl & Uncatalogued & True & Ia & 0.027 & 16.6 & 0.14 & $8.59_{-0.92}^{+0.33}$ & $<0.84$ \\
ASASSN-13dm & 2MASX J03021111+1555387 & True & Ia & 0.017 & 15.6 & 0.36 & $10.29_{-0.07}^{+0.07}$ & $-0.93_{-0.02}^{+0.03}$ \\
ASASSN-14ad & KUG 1237+183 & True & Ia & 0.0264 & 16.9 & 0.05 & $9.71_{-0.47}^{+0.18}$ & $-0.23_{-0.24}^{+0.09}$ \\
ASASSN-14ar & IC 0527 & True & Ia-91bg & 0.02298 & 16.0 & 0.06 & $10.88_{-0.44}^{+0.21}$ & $ 0.20_{-4.20}^{+0.11}$ \\
ASASSN-14as & MGC +06-29-001 & True & Ia & 0.03744 & 16.9 & 0.04 & $10.30_{-0.10}^{+0.09}$ & $-1.30_{-2.70}^{+0.98}$ \\
ASASSN-14ax & SDSS J171000.69+270619.5 & True & Ia & 0.033 & 16.4 & 0.13 & $9.00_{-0.10}^{+0.10}$ & $-0.58_{-0.19}^{+0.04}$ \\
ASASSN-14ay & 2MASX J15570268+3725001 & True & Ia & 0.030869 & 16.3 & 0.06 & $10.19_{-0.13}^{+0.24}$ & $-1.70_{-2.30}^{+0.51}$ \\
ASASSN-14ba & SDSS J102131.91+082419.8 & True & Ia-91T & 0.032668 & 16.8 & 0.08 & $9.19_{-0.08}^{+0.08}$ & $-1.07_{-1.70}^{+0.01}$ \\
ASASSN-14bb & 2MASX J12141125+3839400 & True & Ia & 0.023 & 16.1 & 0.04 & $9.72_{-0.09}^{+0.08}$ & $-1.48_{-0.01}^{+0.00}$ \\
ASASSN-14bd & IC 0831 & True & Ia-91bg & 0.021405 & 16.9 & 0.03 & $11.00_{-0.28}^{+0.04}$ & $-2.19_{-1.81}^{+0.06}$ \\
\hline
\end{tabular}
\medskip
\raggedright
\noindent This table is available in its entirety in a machine readable form. A portion is shown here for guidance regarding its form and content.\\
$^{a}$ This column denotes whether or not this SN/host galaxy pair was used in the analysis.\\
$^{b}$ Peak magnitudes measured from ASAS-SN $V$-band data when possible. For instances when this was not possible, peak magnitudes from D. W. Bishop's Bright Supernova website were adopted and may be from different filters.\\
$^{c}$ The lower limits on the SFR generally not well constrained; we truncate the limits at $10^{-4}\msun$ yr$^{-1}$.\\
\label{tab:host_table}
\end{minipage}
\end{table}

%% file: tab2_sample.tex
\begin{table}
\begin{minipage}{\textwidth}
\caption{Photometric data.$^{a}$\hfill}\begin{tabular}{llllllllc}
\hline
SN name & Host name & $m_{\rm NUV}$ & $m_{\rm SDSS~u}$ & $m_{\rm SDSS~g}$ & $m_{\rm SDSS~r}$ & $m_{\rm SDSS~i}$ & $m_{\rm SDSS~z}$ & Opt. Survey$^{b}$ \\
\hline
ASASSN-13an & 2MASX J13453653-0719350 & $16.8\pm0.1$ & -- & -- & -- & -- & -- & PS1\\
ASASSN-13ar & VV 478 & $17.7\pm0.1$ & $16.9\pm0.1$ & $15.2\pm0.1$ & $14.6\pm0.1$ & $14.3\pm0.1$ & $14.1\pm0.1$ & PS1\\
ASASSN-13av & NGC 7068 & -- & $16.3\pm0.1$ & $14.6\pm0.1$ & $13.7\pm0.1$ & $13.3\pm0.1$ & $12.9\pm0.1$ & SDSS\\
ASASSN-13aw & CGCG 252-043 & $16.8\pm0.1$ & -- & -- & -- & -- & -- & PS1\\
ASASSN-13bb & UGC 01395 & $16.8\pm0.1$ & $15.6\pm0.1$ & $14.1\pm0.1$ & $13.3\pm0.1$ & $12.9\pm0.1$ & $12.6\pm0.1$ & SDSS\\
ASASSN-13cc & NGC 1954 & -- & -- & -- & -- & -- & -- & --\\
ASASSN-13ch & CGCG 023-030 & $17.8\pm0.1$ & $16.9\pm0.1$ & $16.0\pm0.1$ & $15.7\pm0.1$ & $15.5\pm0.1$ & $15.3\pm0.1$ & SDSS\\
ASASSN-13cj & CGCG 051-075 & $20.3\pm0.1$ & $17.2\pm0.1$ & $15.5\pm0.1$ & $14.7\pm0.1$ & $14.3\pm0.1$ & $14.0\pm0.1$ & PS1\\
ASASSN-13cp & ARK 477 & $17.8\pm0.1$ & $16.1\pm0.1$ & $14.3\pm0.1$ & $13.5\pm0.1$ & $13.1\pm0.1$ & $12.7\pm0.1$ & SDSS\\
ASASSN-13cu & VIII Zw 035 & $17.4\pm0.1$ & $16.6\pm0.1$ & $15.4\pm0.1$ & $14.8\pm0.1$ & $14.6\pm0.1$ & $14.3\pm0.1$ & SDSS\\
ASASSN-13dd & NGC 2765 & $17.9\pm0.1$ & $14.5\pm0.1$ & $12.7\pm0.1$ & $11.9\pm0.1$ & $11.5\pm0.1$ & $11.2\pm0.1$ & SDSS\\
ASASSN-13dl & Uncatalogued & -- & -- & -- & -- & -- & -- & PS1\\
ASASSN-13dm & 2MASX J03021111+1555387 & $19.2\pm0.1$ & -- & -- & -- & -- & -- & PS1\\
ASASSN-14ad & KUG 1237+183 & $17.8\pm0.1$ & $17.5\pm0.1$ & $16.5\pm0.1$ & $16.1\pm0.1$ & $16.0\pm0.1$ & $15.9\pm0.1$ & SDSS\\
ASASSN-14ar & IC 0527 & $16.3\pm0.1$ & $18.6\pm0.1$ & $16.9\pm0.1$ & $17.1\pm0.1$ & $15.4\pm0.1$ & $15.4\pm0.1$ & --\\
ASASSN-14as & MGC +06-29-001 & -- & $17.1\pm0.1$ & $15.7\pm0.1$ & $15.0\pm0.1$ & $14.6\pm0.1$ & $14.3\pm0.1$ & SDSS\\
ASASSN-14ax & SDSS J171000.69+270619.5 & $19.4\pm0.1$ & $18.7\pm0.1$ & $17.9\pm0.1$ & $17.6\pm0.1$ & $17.4\pm0.1$ & $17.2\pm0.1$ & SDSS\\
ASASSN-14ay & 2MASX J15570268+3725001 & $20.6\pm0.3$ & $18.0\pm0.1$ & $16.3\pm0.1$ & $15.5\pm0.1$ & $15.2\pm0.1$ & $14.9\pm0.1$ & SDSS\\
ASASSN-14ba & SDSS J102131.91+082419.8 & $19.6\pm0.1$ & $18.7\pm0.1$ & $17.3\pm0.1$ & $16.7\pm0.1$ & $16.4\pm0.1$ & $16.1\pm0.1$ & SDSS\\
ASASSN-14bb & 2MASX J12141125+3839400 & $19.9\pm0.1$ & $17.9\pm0.1$ & $16.5\pm0.1$ & $15.9\pm0.1$ & $15.5\pm0.1$ & $15.3\pm0.1$ & SDSS\\
ASASSN-14bd & IC 0831 & $19.6\pm0.1$ & $16.4\pm0.1$ & $14.6\pm0.1$ & $13.7\pm0.1$ & $13.3\pm0.1$ & $12.9\pm0.1$ & SDSS\\
\hline
\end{tabular}
\medskip
\raggedright
\noindent This table is available in its entirety in a machine readable form. A portion is shown here for guidance regarding its form and content.\\
$^{a}$ The full table contains additional columns for PS1, 2MASS, and WISE magnitudes.\\
$^{b}$ This column denotes which survey produced the optical magnitudes used in the modeling (if any).\\
\label{tab:phot_table}
\end{minipage}
\end{table}

%% file: ms.bbl
\begin{thebibliography}{131}
\expandafter\ifx\csname natexlab\endcsname\relax\def\natexlab#1{#1}\fi

\bibitem[{{Adams} {et~al}\mbox{.}(2013){Adams}, {Kochanek}, {Beacom}, {Vagins},
  \& {Stanek}}]{Adams13}
{Adams} S.~M., {Kochanek} C.~S., {Beacom} J.~F., {Vagins} M.~R., {Stanek}
  K.~Z., 2013, \apj, 778, 164

\bibitem[{{Albareti} {et~al}\mbox{.}(2017){Albareti}, {Allende Prieto},
  {Almeida}, {Anders}, {Anderson}, {Andrews}, {Arag{\'o}n-Salamanca},
  {Argudo-Fern{\'a}ndez}, {Armengaud}, {Aubourg}, \& et~al.}]{Albareti17}
{Albareti} F.~D. {et~al.}, 2017, \apjs, 233, 25

\bibitem[{{Aldering} {et~al}\mbox{.}(2002){Aldering}, {Adam}, {Antilogus},
  {Astier}, {Bacon}, {Bongard}, {Bonnaud}, {Copin}, {Hardin}, {Henault},
  {Howell}, {Lemonnier}, {Levy}, {Loken}, {Nugent}, {Pain}, {Pecontal},
  {Pecontal}, {Perlmutter}, {Quimby}, {Schahmaneche}, {Smadja}, \&
  {Wood-Vasey}}]{Aldering02}
{Aldering} G. {et~al.}, 2002, in \procspie, Vol. 4836, Survey and Other
  Telescope Technologies and Discoveries, {Tyson} J.~A., {Wolff} S., eds., pp.
  61--72

\bibitem[{{Anderson} {et~al}\mbox{.}(2015){Anderson}, {James}, {F{\"o}rster},
  {Gonz{\'a}lez-Gait{\'a}n}, {Habergham}, {Hamuy}, \& {Lyman}}]{Anderson15}
{Anderson} J.~P., {James} P.~A., {F{\"o}rster} F., {Gonz{\'a}lez-Gait{\'a}n}
  S., {Habergham} S.~M., {Hamuy} M., {Lyman} J.~D., 2015, \mnras, 448, 732

\bibitem[{{Astier} {et~al}\mbox{.}(2006){Astier}, {Guy}, {Regnault}, {Pain},
  {Aubourg}, {Balam}, {Basa}, {Carlberg}, {Fabbro}, {Fouchez}, {Hook},
  {Howell}, {Lafoux}, {Neill}, {Palanque-Delabrouille}, {Perrett}, {Pritchet},
  {Rich}, {Sullivan}, {Taillet}, {Aldering}, {Antilogus}, {Arsenijevic},
  {Balland}, {Baumont}, {Bronder}, {Courtois}, {Ellis}, {Filiol}, {Gon{\c
  c}alves}, {Goobar}, {Guide}, {Hardin}, {Lusset}, {Lidman}, {McMahon},
  {Mouchet}, {Mourao}, {Perlmutter}, {Ripoche}, {Tao}, \& {Walton}}]{Astier06}
{Astier} P. {et~al.}, 2006, \aap, 447, 31

\bibitem[{{Baldry} {et~al}\mbox{.}(2012){Baldry}, {Driver}, {Loveday},
  {Taylor}, {Kelvin}, {Liske}, {Norberg}, {Robotham}, {Brough}, {Hopkins},
  {Bamford}, {Peacock}, {Bland-Hawthorn}, {Conselice}, {Croom}, {Jones},
  {Parkinson}, {Popescu}, {Prescott}, {Sharp}, \& {Tuffs}}]{Baldry12}
{Baldry} I.~K. {et~al.}, 2012, \mnras, 421, 621

\bibitem[{{Behroozi}, {Wechsler} \& {Conroy}(2013){Behroozi}, {Wechsler}, \&
  {Conroy}}]{Behroozi13}
{Behroozi} P.~S., {Wechsler} R.~H., {Conroy} C., 2013, \apj, 770, 57

\bibitem[{{Bell} {et~al}\mbox{.}(2003){Bell}, {McIntosh}, {Katz}, \&
  {Weinberg}}]{Bell03}
{Bell} E.~F., {McIntosh} D.~H., {Katz} N., {Weinberg} M.~D., 2003, \apjs, 149,
  289

\bibitem[{{Boogaard} {et~al}\mbox{.}(2018){Boogaard}, {Brinchmann},
  {Bouch{\'e}}, {Paalvast}, {Bacon}, {Bouwens}, {Contini}, {Gunawardhana},
  {Inami}, {Marino}, {Maseda}, {Mitchell}, {Nanayakkara}, {Richard}, {Schaye},
  {Schreiber}, {Tacchella}, {Wisotzki}, \& {Zabl}}]{Boogaard18}
{Boogaard} L.~A. {et~al.}, 2018, ArXiv e-prints

\bibitem[{{Bose} {et~al}\mbox{.}(2018{\natexlab{a}}){Bose}, {Dong}, {Kochanek},
  {Pastorello}, {Katz}, {Bersier}, {Andrews}, {Prieto}, {Stanek}, {Shappee},
  {Smith}, {Kollmeier}, {Benetti}, {Cappellaro}, {Chen}, {Elias-Rosa}, {Milne},
  {Morales-Garoffolo}, {Tartaglia}, {Tomasella}, {Bilinski}, {Brimacombe},
  {Frank}, {Holoien}, {Kilpatrick}, {Kiyota}, {Madore}, \& {Rich}}]{Bose18}
{Bose} S. {et~al.}, 2018{\natexlab{a}}, \apj, 862, 107

\bibitem[{{Bose} {et~al}\mbox{.}(2018{\natexlab{b}}){Bose}, {Dong},
  {Pastorello}, {Filippenko}, {Kochanek}, {Mauerhan}, {Romero-Ca{\~n}izales},
  {Brink}, {Chen}, {Prieto}, {Post}, {Ashall}, {Grupe}, {Tomasella}, {Benetti},
  {Shappee}, {Stanek}, {Cai}, {Falco}, {Lundqvist}, {Mattila}, {Mutel},
  {Ochner}, {Pooley}, {Stritzinger}, {Villanueva}, {Zheng}, {Beswick}, {Brown},
  {Cappellaro}, {Davis}, {Fraser}, {de Jaeger}, {Elias-Rosa}, {Gall}, {Gaudi},
  {Herczeg}, {Hestenes}, {Holoien}, {Hosseinzadeh}, {Hsiao}, {Hu}, {Jaejin},
  {Jeffers}, {Koff}, {Kumar}, {Kurtenkov}, {Lau}, {Prentice}, {Reynolds},
  {Rudy}, {Shahbandeh}, {Somero}, {Stassun}, {Thompson}, {Valenti}, {Woo}, \&
  {Yunus}}]{Bose17}
{Bose} S. {et~al.}, 2018{\natexlab{b}}, \apj, 853, 57

\bibitem[{{Brandt} {et~al}\mbox{.}(2010){Brandt}, {Tojeiro}, {Aubourg},
  {Heavens}, {Jimenez}, \& {Strauss}}]{Brandt10}
{Brandt} T.~D., {Tojeiro} R., {Aubourg} {\'E}., {Heavens} A., {Jimenez} R.,
  {Strauss} M.~A., 2010, \aj, 140, 804

\bibitem[{{Brinchmann} {et~al}\mbox{.}(2004){Brinchmann}, {Charlot}, {White},
  {Tremonti}, {Kauffmann}, {Heckman}, \& {Brinkmann}}]{Brinchmann04}
{Brinchmann} J., {Charlot} S., {White} S.~D.~M., {Tremonti} C., {Kauffmann} G.,
  {Heckman} T., {Brinkmann} J., 2004, \mnras, 351, 1151

\bibitem[{{Brown} {et~al}\mbox{.}(2017){Brown}, {Holoien}, {Auchettl},
  {Stanek}, {Kochanek}, {Shappee}, {Prieto}, \& {Grupe}}]{Brown16_14li}
{Brown} J.~S., {Holoien} T.~W.-S., {Auchettl} K., {Stanek} K.~Z., {Kochanek}
  C.~S., {Shappee} B.~J., {Prieto} J.~L., {Grupe} D., 2017, \mnras, 466, 4904

\bibitem[{{Brown} {et~al}\mbox{.}(2016){Brown}, {Shappee}, {Holoien}, {Stanek},
  {Kochanek}, \& {Prieto}}]{Brown16_14ae}
{Brown} J.~S., {Shappee} B.~J., {Holoien} T.~W.-S., {Stanek} K.~Z., {Kochanek}
  C.~S., {Prieto} J.~L., 2016, \mnras, 462, 3993

\bibitem[{{Bruzual} \& {Charlot}(2003)}]{Bruzual03}
{Bruzual} G., {Charlot} S., 2003, \mnras, 344, 1000

\bibitem[{{Burns} {et~al}\mbox{.}(2011){Burns}, {Stritzinger}, {Phillips},
  {Kattner}, {Persson}, {Madore}, {Freedman}, {Boldt}, {Campillay},
  {Contreras}, {Folatelli}, {Gonzalez}, {Krzeminski}, {Morrell}, {Salgado}, \&
  {Suntzeff}}]{Burns11}
{Burns} C.~R. {et~al.}, 2011, \aj, 141, 19

\bibitem[{{Cardelli}, {Clayton} \& {Mathis}(1989){Cardelli}, {Clayton}, \&
  {Mathis}}]{Cardelli89}
{Cardelli} J.~A., {Clayton} G.~C., {Mathis} J.~S., 1989, \apj, 345, 245

\bibitem[{{Chambers} {et~al}\mbox{.}(2016){Chambers}, {Magnier}, {Metcalfe},
  {Flewelling}, {Huber}, {Waters}, {Denneau}, {Draper}, {Farrow}, {Finkbeiner},
  {Holmberg}, {Koppenhoefer}, {Price}, {Saglia}, {Schlafly}, {Smartt},
  {Sweeney}, {Wainscoat}, {Burgett}, {Grav}, {Heasley}, {Hodapp}, {Jedicke},
  {Kaiser}, {Kudritzki}, {Luppino}, {Lupton}, {Monet}, {Morgan}, {Onaka},
  {Stubbs}, {Tonry}, {Banados}, {Bell}, {Bender}, {Bernard}, {Botticella},
  {Casertano}, {Chastel}, {Chen}, {Chen}, {Cole}, {Deacon}, {Frenk},
  {Fitzsimmons}, {Gezari}, {Goessl}, {Goggia}, {Goldman}, {Grebel}, {Hambly},
  {Hasinger}, {Heavens}, {Heckman}, {Henderson}, {Henning}, {Holman}, {Hopp},
  {Ip}, {Isani}, {Keyes}, {Koekemoer}, {Kotak}, {Long}, {Lucey}, {Liu},
  {Martin}, {McLean}, {Morganson}, {Murphy}, {Nieto-Santisteban}, {Norberg},
  {Peacock}, {Pier}, {Postman}, {Primak}, {Rae}, {Rest}, {Riess}, {Riffeser},
  {Rix}, {Roser}, {Schilbach}, {Schultz}, {Scolnic}, {Szalay}, {Seitz},
  {Shiao}, {Small}, {Smith}, {Soderblom}, {Taylor}, {Thakar}, {Thiel},
  {Thilker}, {Urata}, {Valenti}, {Walter}, {Watters}, {Werner}, {White},
  {Wood-Vasey}, \& {Wyse}}]{Chambers16}
{Chambers} K.~C. {et~al.}, 2016, ArXiv e-prints

\bibitem[{{Childress} {et~al}\mbox{.}(2013{\natexlab{a}}){Childress},
  {Aldering}, {Antilogus}, {Aragon}, {Bailey}, {Baltay}, {Bongard}, {Buton},
  {Canto}, {Cellier-Holzem}, {Chotard}, {Copin}, {Fakhouri}, {Gangler}, {Guy},
  {Hsiao}, {Kerschhaggl}, {Kim}, {Kowalski}, {Loken}, {Nugent}, {Paech},
  {Pain}, {Pecontal}, {Pereira}, {Perlmutter}, {Rabinowitz}, {Rigault},
  {Runge}, {Scalzo}, {Smadja}, {Tao}, {Thomas}, {Weaver}, \&
  {Wu}}]{Childress13a}
{Childress} M. {et~al.}, 2013{\natexlab{a}}, \apj, 770, 107

\bibitem[{{Childress} {et~al}\mbox{.}(2013{\natexlab{b}}){Childress},
  {Aldering}, {Antilogus}, {Aragon}, {Bailey}, {Baltay}, {Bongard}, {Buton},
  {Canto}, {Cellier-Holzem}, {Chotard}, {Copin}, {Fakhouri}, {Gangler}, {Guy},
  {Hsiao}, {Kerschhaggl}, {Kim}, {Kowalski}, {Loken}, {Nugent}, {Paech},
  {Pain}, {Pecontal}, {Pereira}, {Perlmutter}, {Rabinowitz}, {Rigault},
  {Runge}, {Scalzo}, {Smadja}, {Tao}, {Thomas}, {Weaver}, \&
  {Wu}}]{Childress13b}
{Childress} M. {et~al.}, 2013{\natexlab{b}}, \apj, 770, 108

\bibitem[{{Chomiuk} {et~al}\mbox{.}(2012){Chomiuk}, {Soderberg}, {Moe},
  {Chevalier}, {Rupen}, {Badenes}, {Margutti}, {Fransson}, {Fong}, \&
  {Dittmann}}]{Chomiuk12}
{Chomiuk} L. {et~al.}, 2012, \apj, 750, 164

\bibitem[{{Dong} {et~al}\mbox{.}(2015){Dong}, {Katz}, {Kushnir}, \&
  {Prieto}}]{Dong15}
{Dong} S., {Katz} B., {Kushnir} D., {Prieto} J.~L., 2015, \mnras, 454, L61

\bibitem[{{Dong} {et~al}\mbox{.}(2016){Dong}, {Shappee}, {Prieto}, {Jha},
  {Stanek}, {Holoien}, {Kochanek}, {Thompson}, {Morrell}, {Thompson}, {Basu},
  {Beacom}, {Bersier}, {Brimacombe}, {Brown}, {Bufano}, {Chen}, {Conseil},
  {Danilet}, {Falco}, {Grupe}, {Kiyota}, {Masi}, {Nicholls}, {Olivares E.},
  {Pignata}, {Pojmanski}, {Simonian}, {Szczygiel}, \& {Wo{\'z}niak}}]{Dong16}
{Dong} S. {et~al.}, 2016, Science, 351, 257

\bibitem[{{Felten}(1976)}]{Felten76}
{Felten} J.~E., 1976, \apj, 207, 700

\bibitem[{{Filippenko}(1997)}]{Filippenko97}
{Filippenko} A.~V., 1997, \araa, 35, 309

\bibitem[{{Flewelling} {et~al}\mbox{.}(2016){Flewelling}, {Magnier},
  {Chambers}, {Heasley}, {Holmberg}, {Huber}, {Sweeney}, {Waters}, {Chen},
  {Farrow}, {Hasinger}, {Henderson}, {Long}, {Metcalfe}, {Nieto-Santisteban},
  {Norberg}, {Saglia}, {Szalay}, {Rest}, {Thakar}, {Tonry}, {Valenti},
  {Werner}, {White}, {Denneau}, {Draper}, {Hodapp}, {Jedicke}, {Kaiser},
  {Kudritzki}, {Price}, {Wainscoat}, {Chastel}, {McClean}, {Postman}, \&
  {Shiao}}]{Flewelling16}
{Flewelling} H.~A. {et~al.}, 2016, ArXiv e-prints

\bibitem[{{Foley} {et~al}\mbox{.}(2018){Foley}, {Scolnic}, {Rest}, {Jha},
  {Pan}, {Riess}, {Challis}, {Chambers}, {Coulter}, {Dettman}, {Foley}, {Fox},
  {Huber}, {Jones}, {Kilpatrick}, {Kirshner}, {Schultz}, {Siebert},
  {Flewelling}, {Gibson}, {Magnier}, {Miller}, {Primak}, {Smartt}, {Smith},
  {Wainscoat}, {Waters}, \& {Willman}}]{Foley18}
{Foley} R.~J. {et~al.}, 2018, \mnras, 475, 193

\bibitem[{{Fossey} {et~al}\mbox{.}(2014){Fossey}, {Cooke}, {Pollack}, {Wilde},
  \& {Wright}}]{Fossey14}
{Fossey} S.~J., {Cooke} B., {Pollack} G., {Wilde} M., {Wright} T., 2014,
  Central Bureau Electronic Telegrams, 3792

\bibitem[{{Frieman} {et~al}\mbox{.}(2008){Frieman}, {Bassett}, {Becker},
  {Choi}, {Cinabro}, {DeJongh}, {Depoy}, {Dilday}, {Doi}, {Garnavich}, {Hogan},
  {Holtzman}, {Im}, {Jha}, {Kessler}, {Konishi}, {Lampeitl}, {Marriner},
  {Marshall}, {McGinnis}, {Miknaitis}, {Nichol}, {Prieto}, {Riess}, {Richmond},
  {Romani}, {Sako}, {Schneider}, {Smith}, {Takanashi}, {Tokita}, {van der
  Heyden}, {Yasuda}, {Zheng}, {Adelman-McCarthy}, {Annis}, {Assef},
  {Barentine}, {Bender}, {Blandford}, {Boroski}, {Bremer}, {Brewington},
  {Collins}, {Crotts}, {Dembicky}, {Eastman}, {Edge}, {Edmondson}, {Elson},
  {Eyler}, {Filippenko}, {Foley}, {Frank}, {Goobar}, {Gueth}, {Gunn},
  {Harvanek}, {Hopp}, {Ihara}, {Ivezi{\'c}}, {Kahn}, {Kaplan}, {Kent},
  {Ketzeback}, {Kleinman}, {Kollatschny}, {Kron}, {Krzesi{\'n}ski}, {Lamenti},
  {Leloudas}, {Lin}, {Long}, {Lucey}, {Lupton}, {Malanushenko}, {Malanushenko},
  {McMillan}, {Mendez}, {Morgan}, {Morokuma}, {Nitta}, {Ostman}, {Pan},
  {Rockosi}, {Romer}, {Ruiz-Lapuente}, {Saurage}, {Schlesinger}, {Snedden},
  {Sollerman}, {Stoughton}, {Stritzinger}, {Subba Rao}, {Tucker}, {Vaisanen},
  {Watson}, {Watters}, {Wheeler}, {Yanny}, \& {York}}]{Frieman08}
{Frieman} J.~A. {et~al.}, 2008, \aj, 135, 338

\bibitem[{{Gallazzi} {et~al}\mbox{.}(2005){Gallazzi}, {Charlot}, {Brinchmann},
  {White}, \& {Tremonti}}]{Gallazzi05}
{Gallazzi} A., {Charlot} S., {Brinchmann} J., {White} S.~D.~M., {Tremonti}
  C.~A., 2005, \mnras, 362, 41

\bibitem[{{Gao} \& {Pritchet}(2013)}]{Gao13}
{Gao} Y., {Pritchet} C., 2013, \aj, 145, 83

\bibitem[{{Gehrels}(1986)}]{Gehrels86}
{Gehrels} N., 1986, \apj, 303, 336

\bibitem[{{Godoy-Rivera} {et~al}\mbox{.}(2017){Godoy-Rivera}, {Stanek},
  {Kochanek}, {Chen}, {Dong}, {Prieto}, {Shappee}, {Jha}, {Foley}, {Pan},
  {Holoien}, {Thompson}, {Grupe}, \& {Beacom}}]{Godoy17}
{Godoy-Rivera} D. {et~al.}, 2017, \mnras, 466, 1428

\bibitem[{{Goto} {et~al}\mbox{.}(2011){Goto}, {Arnouts}, {Inami}, {Matsuhara},
  {Pearson}, {Takeuchi}, {Le Floc'h}, {Takagi}, {Wada}, {Nakagawa}, {Oyabu},
  {Ishihara}, {Mok Lee}, {Jeong}, {Yamauchi}, {Serjeant}, {Sedgwick}, \&
  {Treister}}]{Goto11}
{Goto} T. {et~al.}, 2011, \mnras, 410, 573

\bibitem[{{Graur} {et~al}\mbox{.}(2017){Graur}, {Bianco}, {Huang}, {Modjaz},
  {Shivvers}, {Filippenko}, {Li}, \& {Eldridge}}]{Graur17}
{Graur} O., {Bianco} F.~B., {Huang} S., {Modjaz} M., {Shivvers} I.,
  {Filippenko} A.~V., {Li} W., {Eldridge} J.~J., 2017, \apj, 837, 120

\bibitem[{{Graur}, {Bianco} \& {Modjaz}(2015){Graur}, {Bianco}, \&
  {Modjaz}}]{Graur15}
{Graur} O., {Bianco} F.~B., {Modjaz} M., 2015, \mnras, 450, 905

\bibitem[{{Graur} \& {Maoz}(2013)}]{Graur13}
{Graur} O., {Maoz} D., 2013, \mnras, 430, 1746

\bibitem[{{Gupta} {et~al}\mbox{.}(2011){Gupta}, {D'Andrea}, {Sako}, {Conroy},
  {Smith}, {Bassett}, {Frieman}, {Garnavich}, {Jha}, {Kessler}, {Lampeitl},
  {Marriner}, {Nichol}, \& {Schneider}}]{Gupta11}
{Gupta} R.~R. {et~al.}, 2011, \apj, 740, 92

\bibitem[{{Guy} {et~al}\mbox{.}(2010){Guy}, {Sullivan}, {Conley}, {Regnault},
  {Astier}, {Balland}, {Basa}, {Carlberg}, {Fouchez}, {Hardin}, {Hook},
  {Howell}, {Pain}, {Palanque-Delabrouille}, {Perrett}, {Pritchet}, {Rich},
  {Ruhlmann-Kleider}, {Balam}, {Baumont}, {Ellis}, {Fabbro}, {Fakhouri},
  {Fourmanoit}, {Gonz{\'a}lez-Gait{\'a}n}, {Graham}, {Hsiao}, {Kronborg},
  {Lidman}, {Mourao}, {Perlmutter}, {Ripoche}, {Suzuki}, \& {Walker}}]{Guy10}
{Guy} J. {et~al.}, 2010, \aap, 523, A7

\bibitem[{{Hamuy} {et~al}\mbox{.}(1995){Hamuy}, {Phillips}, {Maza}, {Suntzeff},
  {Schommer}, \& {Aviles}}]{Hamuy95}
{Hamuy} M., {Phillips} M.~M., {Maza} J., {Suntzeff} N.~B., {Schommer} R.~A.,
  {Aviles} R., 1995, \aj, 109, 1

\bibitem[{{Hamuy} {et~al}\mbox{.}(2000){Hamuy}, {Trager}, {Pinto}, {Phillips},
  {Schommer}, {Ivanov}, \& {Suntzeff}}]{Hamuy00}
{Hamuy} M., {Trager} S.~C., {Pinto} P.~A., {Phillips} M.~M., {Schommer} R.~A.,
  {Ivanov} V., {Suntzeff} N.~B., 2000, \aj, 120, 1479

\bibitem[{{Herczeg} {et~al}\mbox{.}(2016){Herczeg}, {Dong}, {Shappee}, {Chen},
  {Hillenbrand}, {Jose}, {Kochanek}, {Prieto}, {Stanek}, {Kaplan}, {Holoien},
  {Mairs}, {Johnstone}, {Gully-Santiago}, {Zhu}, {Smith}, {Bersier}, {Mulders},
  {Filippenko}, {Ayani}, {Brimacombe}, {Brown}, {Connelley}, {Harmanen},
  {Itoh}, {Kawabata}, {Maehara}, {Takata}, {Yuk}, \& {Zheng}}]{Herczeg16}
{Herczeg} G.~J. {et~al.}, 2016, \apj, 831, 133

\bibitem[{{Heringer} {et~al}\mbox{.}(2017){Heringer}, {Pritchet}, {Kezwer},
  {Graham}, {Sand}, \& {Bildfell}}]{Heringer17}
{Heringer} E., {Pritchet} C., {Kezwer} J., {Graham} M.~L., {Sand} D.,
  {Bildfell} C., 2017, \apj, 834, 15

\bibitem[{{Hogg} {et~al}\mbox{.}(2002){Hogg}, {Baldry}, {Blanton}, \&
  {Eisenstein}}]{Hogg02}
{Hogg} D.~W., {Baldry} I.~K., {Blanton} M.~R., {Eisenstein} D.~J., 2002, ArXiv
  Astrophysics e-prints

\bibitem[{{Holoien} {et~al}\mbox{.}(2017{\natexlab{a}}){Holoien}, {Brown},
  {Stanek}, {Kochanek}, {Shappee}, {Prieto}, {Dong}, {Brimacombe}, {Bishop},
  {Basu}, {Beacom}, {Bersier}, {Chen}, {Danilet}, {Falco}, {Godoy-Rivera},
  {Goss}, {Pojmanski}, {Simonian}, {Skowron}, {Thompson}, {Wo{\'z}niak},
  {{\'A}vila}, {Bock}, {Carballo}, {Conseil}, {Contreras}, {Cruz},
  {And{\'u}jar}, {Guo}, {Hsiao}, {Kiyota}, {Koff}, {Krannich}, {Madore},
  {Marples}, {Masi}, {Morrell}, {Monard}, {Munoz-Mateos}, {Nicholls},
  {Nicolas}, {Wagner}, \& {Wiethoff}}]{snCat2}
{Holoien} T.~W.-S. {et~al.}, 2017{\natexlab{a}}, \mnras, 467, 1098

\bibitem[{{Holoien} {et~al}\mbox{.}(2017{\natexlab{b}}){Holoien}, {Brown},
  {Stanek}, {Kochanek}, {Shappee}, {Prieto}, {Dong}, {Brimacombe}, {Bishop},
  {Bose}, {Beacom}, {Bersier}, {Chen}, {Chomiuk}, {Falco}, {Godoy-Rivera},
  {Morrell}, {Pojmanski}, {Shields}, {Strader}, {Stritzinger}, {Thompson},
  {Wo{\'z}niak}, {Bock}, {Cacella}, {Conseil}, {Cruz}, {Fernandez}, {Kiyota},
  {Koff}, {Krannich}, {Marples}, {Masi}, {Monard}, {Nicholls}, {Nicolas},
  {Post}, {Stone}, \& {Wiethoff}}]{snCat3}
{Holoien} T.~W.-S. {et~al.}, 2017{\natexlab{b}}, \mnras, 471, 4966

\bibitem[{{Holoien} {et~al}\mbox{.}(2016{\natexlab{a}}){Holoien}, {Kochanek},
  {Prieto}, {Grupe}, {Chen}, {Godoy-Rivera}, {Stanek}, {Shappee}, {Dong},
  {Brown}, {Basu}, {Beacom}, {Bersier}, {Brimacombe}, {Carlson}, {Falco},
  {Johnston}, {Madore}, {Pojmanski}, \& {Seibert}}]{Holoien16_15oi}
{Holoien} T.~W.-S. {et~al.}, 2016{\natexlab{a}},
  http://arxiv.org/abs/1602.01088

\bibitem[{{Holoien} {et~al}\mbox{.}(2016{\natexlab{b}}){Holoien}, {Kochanek},
  {Prieto}, {Stanek}, {Dong}, {Shappee}, {Grupe}, {Brown}, {Basu}, {Beacom},
  {Bersier}, {Brimacombe}, {Danilet}, {Falco}, {Guo}, {Jose}, {Herczeg},
  {Long}, {Pojmanski}, {Simonian}, {Szczygie{\l}}, {Thompson}, {Thorstensen},
  {Wagner}, \& {Wo{\'z}niak}}]{Holoien16_14li}
{Holoien} T.~W.-S. {et~al.}, 2016{\natexlab{b}}, \mnras, 455, 2918

\bibitem[{{Holoien} {et~al}\mbox{.}(2014){Holoien}, {Prieto}, {Bersier},
  {Kochanek}, {Stanek}, {Shappee}, {Grupe}, {Basu}, {Beacom}, {Brimacombe},
  {Brown}, {Davis}, {Jencson}, {Pojmanski}, \& {Szczygie{\l}}}]{Holoien14}
{Holoien} T.~W.-S. {et~al.}, 2014, \mnras, 445, 3263

\bibitem[{{Holoien} {et~al}\mbox{.}(2016{\natexlab{c}}){Holoien}, {Prieto},
  {Pejcha}, {Stanek}, {Kochanek}, {Shappee}, {Grupe}, {Morrell}, {Thorstensen},
  {Basu}, {Beacom}, {Bersier}, {Brimacombe}, {Davis}, {Pojma{\'n}ski}, \&
  {Skowron}}]{Holoien16_acta}
{Holoien} T.~W.-S. {et~al.}, 2016{\natexlab{c}}, \actaa, 66, 219

\bibitem[{{Holoien} {et~al}\mbox{.}(2017{\natexlab{c}}){Holoien}, {Stanek},
  {Kochanek}, {Shappee}, {Prieto}, {Brimacombe}, {Bersier}, {Bishop}, {Dong},
  {Brown}, {Danilet}, {Simonian}, {Basu}, {Beacom}, {Falco}, {Pojmanski},
  {Skowron}, {Wo{\'z}niak}, {{\'A}vila}, {Conseil}, {Contreras}, {Cruz},
  {Fern{\'a}ndez}, {Koff}, {Guo}, {Herczeg}, {Hissong}, {Hsiao}, {Jose},
  {Kiyota}, {Long}, {Monard}, {Nicholls}, {Nicolas}, \& {Wiethoff}}]{snCat1}
{Holoien} T.~W.-S. {et~al.}, 2017{\natexlab{c}}, \mnras, 464, 2672

\bibitem[{{Hopkins} \& {Beacom}(2006)}]{Hopkins06}
{Hopkins} A.~M., {Beacom} J.~F., 2006, \apj, 651, 142

\bibitem[{{Hoyle} \& {Fowler}(1960)}]{Hoyle60}
{Hoyle} F., {Fowler} W.~A., 1960, \apj, 132, 565

\bibitem[{{Hsiao} {et~al}\mbox{.}(2007){Hsiao}, {Conley}, {Howell}, {Sullivan},
  {Pritchet}, {Carlberg}, {Nugent}, \& {Phillips}}]{Hsiao07}
{Hsiao} E.~Y., {Conley} A., {Howell} D.~A., {Sullivan} M., {Pritchet} C.~J.,
  {Carlberg} R.~G., {Nugent} P.~E., {Phillips} M.~M., 2007, \apj, 663, 1187

\bibitem[{{Huchra} \& {Sargent}(1973)}]{Huchra73}
{Huchra} J., {Sargent} W.~L.~W., 1973, \apj, 186, 433

\bibitem[{{Iben} \& {Tutukov}(1984)}]{Iben84}
{Iben}, Jr. I., {Tutukov} A.~V., 1984, \apjs, 54, 335

\bibitem[{{Jarrett} {et~al}\mbox{.}(2003){Jarrett}, {Chester}, {Cutri},
  {Schneider}, \& {Huchra}}]{Jarrett03}
{Jarrett} T.~H., {Chester} T., {Cutri} R., {Schneider} S.~E., {Huchra} J.~P.,
  2003, \aj, 125, 525

\bibitem[{{Johansson} {et~al}\mbox{.}(2013){Johansson}, {Thomas}, {Pforr},
  {Maraston}, {Nichol}, {Smith}, {Lampeitl}, {Beifiori}, {Gupta}, \&
  {Schneider}}]{Johansson13}
{Johansson} J. {et~al.}, 2013, \mnras, 435, 1680

\bibitem[{{Jones}, {Riess} \& {Scolnic}(2015){Jones}, {Riess}, \&
  {Scolnic}}]{Jones15}
{Jones} D.~O., {Riess} A.~G., {Scolnic} D.~M., 2015, \apj, 812, 31

\bibitem[{{Kato} {et~al}\mbox{.}(2014{\natexlab{a}}){Kato}, {Dubovsky},
  {Kudzej}, {Hambsch}, {Miller}, {Ohshima}, {Nakata}, {Kawabata}, {Nishino},
  {Masumoto}, {Mizoguchi}, {Yamanaka}, {Matsumoto}, {Sakai}, {Fukushima},
  {Matsuura}, {Bouno}, {Takenaka}, {Nakagawa}, {Noguchi}, {Iino}, {Pickard},
  {Maeda}, {Henden}, {Kasai}, {Kiyota}, {Akazawa}, {Imamura}, {de Miguel},
  {Maehara}, {Monard}, {Pavlenko}, {Antonyuk}, {Pit}, {Antonyuk}, {Baklanov},
  {Ruiz}, {Richmond}, {Oksanen}, {Harlingten}, {Shugarov}, {Chochol}, {Masi},
  {Nocentini}, {Schmeer}, {Bolt}, {Nelson}, {Ulowetz}, {Sabo}, {Goff}, {Stein},
  {Michel}, {Dvorak}, {Voloshina}, {Metlov}, {Katysheva}, {Neustroev},
  {Sjoberg}, {Littlefield}, {D{\c e}bski}, {Sowicka}, {Klimaszewski},
  {Cury{\l}o}, {Morelle}, {Curtis}, {Iwamatsu}, {Butterworth}, {Andreev},
  {Parakhin}, {Sklyanov}, {Shiokawa}, {Nov{\'a}k}, {Irsmambetova}, {Itoh},
  {Ito}, {Hirosawa}, {Denisenko}, {Kochanek}, {Shappee}, {Stanek}, {Prieto},
  {Itagaki}, {Stubbings}, {Ripero}, {Muyllaert}, \& {Poyner}}]{Kato14a}
{Kato} T. {et~al.}, 2014{\natexlab{a}}, \pasj, 66, 90

\bibitem[{{Kato} {et~al}\mbox{.}(2015){Kato}, {Hambsch}, {Dubovsky}, {Kudzej},
  {Monard}, {Miller}, {Itoh}, {Kiyota}, {Masumoto}, {Fukushima}, {Kinoshita},
  {Maeda}, {Mikami}, {Matsuda}, {Kojiguchi}, {Kawabata}, {Takenaka},
  {Matsumoto}, {de Miguel}, {Maeda}, {Ohshima}, {Isogai}, {Pickard}, {Henden},
  {Kafka}, {Akazawa}, {Otani}, {Ishibashi}, {Ogi}, {Tanabe}, {Imamura},
  {Stein}, {Kasai}, {Vanmunster}, {Starr}, {Oksanen}, {Pavlenko}, {Antonyuk},
  {Antonyuk}, {Sosnovskij}, {Pit}, {Babina}, {Sklyanov}, {Nov{\'a}k}, {Dvorak},
  {Michel}, {Masi}, {Littlefield}, {Ulowetz}, {Shugarov}, {Golysheva},
  {Chochol}, {Krushevska}, {Ruiz}, {Tordai}, {Morelle}, {Sabo}, {Maehara},
  {Richmond}, {Katysheva}, {Hirosawa}, {Goff}, {Dubois}, {Logie}, {Rau},
  {Voloshina}, {Andreev}, {Shiokawa}, {Neustroev}, {Sjoberg}, {Zharikov},
  {James}, {Bolt}, {Crawford}, {Buczynski}, {Cook}, {Kochanek}, {Shappee},
  {Stanek}, {Prieto}, {Denisenko}, {Nishimura}, {Mukai}, {Kaneko}, {Ueda},
  {Stubbings}, {Moriyama}, {Schmeer}, {Muyllaert}, {Shears}, {Modic}, \&
  {Paxson}}]{Kato15}
{Kato} T. {et~al.}, 2015, \pasj, 67, 105

\bibitem[{{Kato} {et~al}\mbox{.}(2014{\natexlab{b}}){Kato}, {Hambsch},
  {Maehara}, {Masi}, {Nocentini}, {Dubovsky}, {Kudzej}, {Imamura}, {Ogi},
  {Tanabe}, {Akazawa}, {Krajci}, {Miller}, {de Miguel}, {Henden}, {Noguchi},
  {Ishibashi}, {Ono}, {Kawabata}, {Kobayashi}, {Sakai}, {Nishino}, {Furukawa},
  {Masumoto}, {Matsumoto}, {Littlefield}, {Ohshima}, {Nakata}, {Honda},
  {Kinugasa}, {Hashimoto}, {Stein}, {Pickard}, {Kiyota}, {Pavlenko},
  {Antonyuk}, {Baklanov}, {Antonyuk}, {Samsonov}, {Pit}, {Sosnovskij},
  {Oksanen}, {Harlingten}, {Tyysk{\"a}}, {Monard}, {Shugarov}, {Chochol},
  {Kasai}, {Maeda}, {Hirosawa}, {Itoh}, {Sabo}, {Ulowetz}, {Morelle}, {Michel},
  {Su{\'a}rez}, {James}, {Dvorak}, {Voloshina}, {Richmond}, {Staels}, {Boyd},
  {Andreev}, {Parakhin}, {Katysheva}, {Miyashita}, {Nakajima}, {Bolt},
  {Padovan}, {Nelson}, {Starkey}, {Buczynski}, {Starr}, {Goff}, {Denisenko},
  {Kochanek}, {Shappee}, {Stanek}, {Prieto}, {Itagaki}, {Kaneko}, {Stubbings},
  {Muyllaert}, {Shears}, {Schmeer}, {Poyner}, \&
  {Rodr{\'{\i}}guez-Marco}}]{Kato14b}
{Kato} T. {et~al.}, 2014{\natexlab{b}}, \pasj, 66, 30

\bibitem[{{Kato} {et~al}\mbox{.}(2016){Kato}, {Hambsch}, {Monard},
  {Vanmunster}, {Maeda}, {Miller}, {Itoh}, {Kiyota}, {Isogai}, {Kimura},
  {Imada}, {Tordai}, {Akazawa}, {Tanabe}, {Otani}, {Ogi}, {Ando}, {Takigawa},
  {Dubovsky}, {Kudzej}, {Shugarov}, {Katysheva}, {Golysheva}, {Gladilina},
  {Chochol}, {Starr}, {Kasai}, {Pickard}, {de Miguel}, {Kojiguchi}, {Sugiura},
  {Fukushima}, {Yamada}, {Uto}, {Kamibetsunawa}, {Tatsumi}, {Takeda},
  {Matsumoto}, {Cook}, {Pavlenko}, {Babina}, {Pit}, {Antonyuk}, {Antonyuk},
  {Sosnovskij}, {Baklanov}, {Kafka}, {Stein}, {Voloshina}, {Ruiz}, {Sabo},
  {Dvorak}, {Stone}, {Andreev}, {Antipin}, {Zubareva}, {Zaostrojnykh},
  {Richmond}, {Shears}, {Dubois}, {Logie}, {Rau}, {Vanaverbeke}, {Simon},
  {Oksanen}, {Goff}, {Bolt}, {D{\c e}bski}, {Kochanek}, {Shappee}, {Stanek},
  {Prieto}, {Stubbings}, {Muyllaert}, {Hiraga}, {Horie}, {Schmeer}, \&
  {Hirosawa}}]{Kato16}
{Kato} T. {et~al.}, 2016, \pasj, 68, 65

\bibitem[{{Kauffmann} {et~al}\mbox{.}(2003){Kauffmann}, {Heckman}, {White},
  {Charlot}, {Tremonti}, {Brinchmann}, {Bruzual}, {Peng}, {Seibert},
  {Bernardi}, {Blanton}, {Brinkmann}, {Castander}, {Cs{\'a}bai}, {Fukugita},
  {Ivezic}, {Munn}, {Nichol}, {Padmanabhan}, {Thakar}, {Weinberg}, \&
  {York}}]{Kauffmann03}
{Kauffmann} G. {et~al.}, 2003, \mnras, 341, 33

\bibitem[{{Kelly} {et~al}\mbox{.}(2010){Kelly}, {Hicken}, {Burke}, {Mandel}, \&
  {Kirshner}}]{Kelly10}
{Kelly} P.~L., {Hicken} M., {Burke} D.~L., {Mandel} K.~S., {Kirshner} R.~P.,
  2010, \apj, 715, 743

\bibitem[{{Khan} {et~al}\mbox{.}(2011{\natexlab{a}}){Khan}, {Prieto},
  {Pojma{\'n}ski}, {Stanek}, {Beacom}, {Szczygiel}, {Pilecki}, {Mogren},
  {Eastman}, {Martini}, \& {Stoll}}]{Khan11a}
{Khan} R. {et~al.}, 2011{\natexlab{a}}, \apj, 726, 106

\bibitem[{{Khan} {et~al}\mbox{.}(2011{\natexlab{b}}){Khan}, {Stanek}, {Stoll},
  \& {Prieto}}]{Khan11b}
{Khan} R., {Stanek} K.~Z., {Stoll} R., {Prieto} J.~L., 2011{\natexlab{b}},
  \apjl, 737, L24

\bibitem[{{Kim}, {Goobar} \& {Perlmutter}(1996){Kim}, {Goobar}, \&
  {Perlmutter}}]{Kim96}
{Kim} A., {Goobar} A., {Perlmutter} S., 1996, \pasp, 108, 190

\bibitem[{{Kistler} {et~al}\mbox{.}(2013){Kistler}, {Stanek}, {Kochanek},
  {Prieto}, \& {Thompson}}]{Kistler13}
{Kistler} M.~D., {Stanek} K.~Z., {Kochanek} C.~S., {Prieto} J.~L., {Thompson}
  T.~A., 2013, \apj, 770, 88

\bibitem[{{Kochanek} {et~al}\mbox{.}(2017){Kochanek}, {Shappee}, {Stanek},
  {Holoien}, {Thompson}, {Prieto}, {Dong}, {Shields}, {Will}, {Britt},
  {Perzanowski}, \& {Pojma{\'n}ski}}]{Kochanek17}
{Kochanek} C.~S. {et~al.}, 2017, \pasp, 129, 104502

\bibitem[{{Kriek} {et~al}\mbox{.}(2009){Kriek}, {van Dokkum}, {Labb{\'e}},
  {Franx}, {Illingworth}, {Marchesini}, \& {Quadri}}]{Kriek09}
{Kriek} M., {van Dokkum} P.~G., {Labb{\'e}} I., {Franx} M., {Illingworth}
  G.~D., {Marchesini} D., {Quadri} R.~F., 2009, \apj, 700, 221

\bibitem[{{Lampeitl} {et~al}\mbox{.}(2010){Lampeitl}, {Smith}, {Nichol},
  {Bassett}, {Cinabro}, {Dilday}, {Foley}, {Frieman}, {Garnavich}, {Goobar},
  {Im}, {Jha}, {Marriner}, {Miquel}, {Nordin}, {{\"O}stman}, {Riess}, {Sako},
  {Schneider}, {Sollerman}, \& {Stritzinger}}]{Lampeitl10}
{Lampeitl} H. {et~al.}, 2010, \apj, 722, 566

\bibitem[{{Law} {et~al}\mbox{.}(2009){Law}, {Kulkarni}, {Dekany}, {Ofek},
  {Quimby}, {Nugent}, {Surace}, {Grillmair}, {Bloom}, {Kasliwal}, {Bildsten},
  {Brown}, {Cenko}, {Ciardi}, {Croner}, {Djorgovski}, {van Eyken},
  {Filippenko}, {Fox}, {Gal-Yam}, {Hale}, {Hamam}, {Helou}, {Henning},
  {Howell}, {Jacobsen}, {Laher}, {Mattingly}, {McKenna}, {Pickles},
  {Poznanski}, {Rahmer}, {Rau}, {Rosing}, {Shara}, {Smith}, {Starr},
  {Sullivan}, {Velur}, {Walters}, \& {Zolkower}}]{Law09}
{Law} N.~M. {et~al.}, 2009, \pasp, 121, 1395

\bibitem[{{Li} {et~al}\mbox{.}(2011{\natexlab{a}}){Li}, {Chornock}, {Leaman},
  {Filippenko}, {Poznanski}, {Wang}, {Ganeshalingam}, \& {Mannucci}}]{Loss3}
{Li} W., {Chornock} R., {Leaman} J., {Filippenko} A.~V., {Poznanski} D., {Wang}
  X., {Ganeshalingam} M., {Mannucci} F., 2011{\natexlab{a}}, \mnras, 412, 1473

\bibitem[{{Li} {et~al}\mbox{.}(2003){Li}, {Filippenko}, {Chornock}, {Berger},
  {Berlind}, {Calkins}, {Challis}, {Fassnacht}, {Jha}, {Kirshner}, {Matheson},
  {Sargent}, {Simcoe}, {Smith}, \& {Squires}}]{Li03}
{Li} W. {et~al.}, 2003, \pasp, 115, 453

\bibitem[{{Li} {et~al}\mbox{.}(2011{\natexlab{b}}){Li}, {Leaman}, {Chornock},
  {Filippenko}, {Poznanski}, {Ganeshalingam}, {Wang}, {Modjaz}, {Jha}, {Foley},
  \& {Smith}}]{Loss2}
{Li} W. {et~al.}, 2011{\natexlab{b}}, \mnras, 412, 1441

\bibitem[{{Li} {et~al}\mbox{.}(2000){Li}, {Filippenko}, {Treffers}, {Friedman},
  {Halderson}, {Johnson}, {King}, {Modjaz}, {Papenkova}, {Sato}, \&
  {Shefler}}]{Li00}
{Li} W.~D. {et~al.}, 2000, in American Institute of Physics Conference Series,
  Vol. 522, American Institute of Physics Conference Series, {Holt} S.~S.,
  {Zhang} W.~W., eds., pp. 103--106

\bibitem[{{Lonsdale}, {Farrah} \& {Smith}(2006){Lonsdale}, {Farrah}, \&
  {Smith}}]{Lonsdale06}
{Lonsdale} C.~J., {Farrah} D., {Smith} H.~E., 2006, {Ultraluminous Infrared
  Galaxies}, {Mason} J.~W., ed., p. 285

\bibitem[{{Madau} \& {Dickinson}(2014)}]{Madau14}
{Madau} P., {Dickinson} M., 2014, \araa, 52, 415

\bibitem[{{Madau} {et~al}\mbox{.}(1996){Madau}, {Ferguson}, {Dickinson},
  {Giavalisco}, {Steidel}, \& {Fruchter}}]{Madau96}
{Madau} P., {Ferguson} H.~C., {Dickinson} M.~E., {Giavalisco} M., {Steidel}
  C.~C., {Fruchter} A., 1996, \mnras, 283, 1388

\bibitem[{{Mannucci} {et~al}\mbox{.}(2005){Mannucci}, {Della Valle}, {Panagia},
  {Cappellaro}, {Cresci}, {Maiolino}, {Petrosian}, \& {Turatto}}]{Mannucci05}
{Mannucci} F., {Della Valle} M., {Panagia} N., {Cappellaro} E., {Cresci} G.,
  {Maiolino} R., {Petrosian} A., {Turatto} M., 2005, \aap, 433, 807

\bibitem[{{Maoz} \& {Mannucci}(2012)}]{MaozMannucci12}
{Maoz} D., {Mannucci} F., 2012, \pasa, 29, 447

\bibitem[{{Maoz}, {Mannucci} \& {Brandt}(2012){Maoz}, {Mannucci}, \&
  {Brandt}}]{Maoz12}
{Maoz} D., {Mannucci} F., {Brandt} T.~D., 2012, \mnras, 426, 3282

\bibitem[{{Maoz} {et~al}\mbox{.}(2011){Maoz}, {Mannucci}, {Li}, {Filippenko},
  {Della Valle}, \& {Panagia}}]{Maoz11}
{Maoz} D., {Mannucci} F., {Li} W., {Filippenko} A.~V., {Della Valle} M.,
  {Panagia} N., 2011, \mnras, 412, 1508

\bibitem[{{Moreno-Raya} {et~al}\mbox{.}(2016){Moreno-Raya},
  {L{\'o}pez-S{\'a}nchez}, {Moll{\'a}}, {Galbany}, {V{\'{\i}}lchez}, \&
  {Carnero}}]{MorenoRaya16}
{Moreno-Raya} M.~E., {L{\'o}pez-S{\'a}nchez} {\'A}.~R., {Moll{\'a}} M.,
  {Galbany} L., {V{\'{\i}}lchez} J.~M., {Carnero} A., 2016, \mnras, 462, 1281

\bibitem[{{Morrissey} {et~al}\mbox{.}(2007){Morrissey}, {Conrow}, {Barlow},
  {Small}, {Seibert}, {Wyder}, {Budav{\'a}ri}, {Arnouts}, {Friedman},
  {Forster}, {Martin}, {Neff}, {Schiminovich}, {Bianchi}, {Donas}, {Heckman},
  {Lee}, {Madore}, {Milliard}, {Rich}, {Szalay}, {Welsh}, \&
  {Yi}}]{Morrissey07}
{Morrissey} P. {et~al.}, 2007, \apjs, 173, 682

\bibitem[{{Neill} {et~al}\mbox{.}(2006){Neill}, {Sullivan}, {Balam},
  {Pritchet}, {Howell}, {Perrett}, {Astier}, {Aubourg}, {Basa}, {Carlberg},
  {Conley}, {Fabbro}, {Fouchez}, {Guy}, {Hook}, {Pain},
  {Palanque-Delabrouille}, {Regnault}, {Rich}, {Taillet}, {Aldering},
  {Antilogus}, {Arsenijevic}, {Balland}, {Baumont}, {Bronder}, {Ellis},
  {Filiol}, {Gon{\c c}alves}, {Hardin}, {Kowalski}, {Lidman}, {Lusset},
  {Mouchet}, {Mourao}, {Perlmutter}, {Ripoche}, {Schlegel}, \& {Tao}}]{Neill06}
{Neill} J.~D. {et~al.}, 2006, \aj, 132, 1126

\bibitem[{{Neill} {et~al}\mbox{.}(2009){Neill}, {Sullivan}, {Howell}, {Conley},
  {Seibert}, {Martin}, {Barlow}, {Foster}, {Friedman}, {Morrissey}, {Neff},
  {Schiminovich}, {Wyder}, {Bianchi}, {Donas}, {Heckman}, {Lee}, {Madore},
  {Milliard}, {Rich}, \& {Szalay}}]{Neill09}
{Neill} J.~D. {et~al.}, 2009, \apj, 707, 1449

\bibitem[{{Nomoto}(1982)}]{Nomoto82}
{Nomoto} K., 1982, \apj, 253, 798

\bibitem[{{Nugent} {et~al}\mbox{.}(2011){Nugent}, {Sullivan}, {Cenko},
  {Thomas}, {Kasen}, {Howell}, {Bersier}, {Bloom}, {Kulkarni}, {Kandrashoff},
  {Filippenko}, {Silverman}, {Marcy}, {Howard}, {Isaacson}, {Maguire},
  {Suzuki}, {Tarlton}, {Pan}, {Bildsten}, {Fulton}, {Parrent}, {Sand},
  {Podsiadlowski}, {Bianco}, {Dilday}, {Graham}, {Lyman}, {James}, {Kasliwal},
  {Law}, {Quimby}, {Hook}, {Walker}, {Mazzali}, {Pian}, {Ofek}, {Gal-Yam}, \&
  {Poznanski}}]{Nugent11}
{Nugent} P.~E. {et~al.}, 2011, \nat, 480, 344

\bibitem[{{O'Donnell}(1994)}]{Odonnell94}
{O'Donnell} J.~E., 1994, \apj, 422, 158

\bibitem[{{Pan} {et~al}\mbox{.}(2014){Pan}, {Sullivan}, {Maguire}, {Hook},
  {Nugent}, {Howell}, {Arcavi}, {Botyanszki}, {Cenko}, {DeRose}, {Fakhouri},
  {Gal-Yam}, {Hsiao}, {Kulkarni}, {Laher}, {Lidman}, {Nordin}, {Walker}, \&
  {Xu}}]{Pan14}
{Pan} Y.-C. {et~al.}, 2014, \mnras, 438, 1391

\bibitem[{{Perlmutter} {et~al}\mbox{.}(1999){Perlmutter}, {Aldering},
  {Goldhaber}, {Knop}, {Nugent}, {Castro}, {Deustua}, {Fabbro}, {Goobar},
  {Groom}, {Hook}, {Kim}, {Kim}, {Lee}, {Nunes}, {Pain}, {Pennypacker},
  {Quimby}, {Lidman}, {Ellis}, {Irwin}, {McMahon}, {Ruiz-Lapuente}, {Walton},
  {Schaefer}, {Boyle}, {Filippenko}, {Matheson}, {Fruchter}, {Panagia},
  {Newberg}, {Couch}, \& {Project}}]{Perlmutter99}
{Perlmutter} S. {et~al.}, 1999, \apj, 517, 565

\bibitem[{{Phillips}(1993)}]{Phillips93}
{Phillips} M.~M., 1993, \apjl, 413, L105

\bibitem[{{Planck Collaboration} {et~al}\mbox{.}(2016){Planck Collaboration},
  {Ade}, {Aghanim}, {Arnaud}, {Ashdown}, {Aumont}, {Baccigalupi}, {Banday},
  {Barreiro}, {Bartlett}, \& et~al.}]{Planck16}
{Planck Collaboration} {et~al.}, 2016, \aap, 594, A13

\bibitem[{{Prieto}, {Stanek} \& {Beacom}(2008){Prieto}, {Stanek}, \&
  {Beacom}}]{Prieto08}
{Prieto} J.~L., {Stanek} K.~Z., {Beacom} J.~F., 2008, \apj, 673, 999

\bibitem[{{Quimby}(2006)}]{Quimby06}
{Quimby} R.~M., 2006, PhD thesis, The University of Texas at Austin

\bibitem[{{Quimby} {et~al}\mbox{.}(2012){Quimby}, {Yuan}, {Akerlof}, {Wheeler},
  \& {Warren}}]{Quimby12}
{Quimby} R.~M., {Yuan} F., {Akerlof} C., {Wheeler} J.~C., {Warren} M.~S., 2012,
  \aj, 144, 177

\bibitem[{{Riess} {et~al}\mbox{.}(1998){Riess}, {Filippenko}, {Challis},
  {Clocchiatti}, {Diercks}, {Garnavich}, {Gilliland}, {Hogan}, {Jha},
  {Kirshner}, {Leibundgut}, {Phillips}, {Reiss}, {Schmidt}, {Schommer},
  {Smith}, {Spyromilio}, {Stubbs}, {Suntzeff}, \& {Tonry}}]{Riess98}
{Riess} A.~G. {et~al.}, 1998, \aj, 116, 1009

\bibitem[{{Riess}, {Press} \& {Kirshner}(1996){Riess}, {Press}, \&
  {Kirshner}}]{Riess96}
{Riess} A.~G., {Press} W.~H., {Kirshner} R.~P., 1996, \apj, 473, 88

\bibitem[{{Rigault} {et~al}\mbox{.}(2015){Rigault}, {Aldering}, {Kowalski},
  {Copin}, {Antilogus}, {Aragon}, {Bailey}, {Baltay}, {Baugh}, {Bongard},
  {Boone}, {Buton}, {Chen}, {Chotard}, {Fakhouri}, {Feindt}, {Fagrelius},
  {Fleury}, {Fouchez}, {Gangler}, {Hayden}, {Kim}, {Leget}, {Lombardo},
  {Nordin}, {Pain}, {Pecontal}, {Pereira}, {Perlmutter}, {Rabinowitz}, {Runge},
  {Rubin}, {Saunders}, {Smadja}, {Sofiatti}, {Suzuki}, {Tao}, \&
  {Weaver}}]{Rigault15}
{Rigault} M. {et~al.}, 2015, \apj, 802, 20

\bibitem[{{Rigault} {et~al}\mbox{.}(2013){Rigault}, {Copin}, {Aldering},
  {Antilogus}, {Aragon}, {Bailey}, {Baltay}, {Bongard}, {Buton}, {Canto},
  {Cellier-Holzem}, {Childress}, {Chotard}, {Fakhouri}, {Feindt}, {Fleury},
  {Gangler}, {Greskovic}, {Guy}, {Kim}, {Kowalski}, {Lombardo}, {Nordin},
  {Nugent}, {Pain}, {P{\'e}contal}, {Pereira}, {Perlmutter}, {Rabinowitz},
  {Runge}, {Saunders}, {Scalzo}, {Smadja}, {Tao}, {Thomas}, \&
  {Weaver}}]{Rigault13}
{Rigault} M. {et~al.}, 2013, \aap, 560, A66

\bibitem[{{Roman} {et~al}\mbox{.}(2017){Roman}, {Hardin}, {Betoule}, {Astier},
  {Balland}, {Ellis}, {Fabbro}, {Guy}, {Hook}, {Howell}, {Lidman}, {Mitra},
  {M{\"o}ller}, {Mour{\~a}o}, {Neveu}, {Palanque-Delabrouille}, {Pritchet},
  {Regnault}, {Ruhlmann-Kleider}, {Saunders}, \& {Sullivan}}]{Roman17}
{Roman} M. {et~al.}, 2017, ArXiv e-prints

\bibitem[{{Scannapieco} \& {Bildsten}(2005)}]{Scannapieco05}
{Scannapieco} E., {Bildsten} L., 2005, \apjl, 629, L85

\bibitem[{{Schechter}(1976)}]{Schechter76}
{Schechter} P., 1976, \apj, 203, 297

\bibitem[{{Schlafly} \& {Finkbeiner}(2011)}]{Schlafly11}
{Schlafly} E.~F., {Finkbeiner} D.~P., 2011, \apj, 737, 103

\bibitem[{{Schmidt}(1968)}]{Schmidt68}
{Schmidt} M., 1968, \apj, 151, 393

\bibitem[{{Schmidt} {et~al}\mbox{.}(2014){Schmidt}, {Prieto}, {Stanek},
  {Shappee}, {Morrell}, {Bardalez Gagliuffi}, {Kochanek}, {Jencson}, {Holoien},
  {Basu}, {Beacom}, {Szczygie{\l}}, {Pojmanski}, {Brimacombe}, {Dubberley},
  {Elphick}, {Foale}, {Hawkins}, {Mullins}, {Rosing}, {Ross}, \&
  {Walker}}]{Schmidt14}
{Schmidt} S.~J. {et~al.}, 2014, \apjl, 781, L24

\bibitem[{{Schmidt} {et~al}\mbox{.}(2016){Schmidt}, {Shappee}, {Gagn{\'e}},
  {Stanek}, {Prieto}, {Holoien}, {Kochanek}, {Chomiuk}, {Dong}, {Seibert}, \&
  {Strader}}]{Schmidt16}
{Schmidt} S.~J. {et~al.}, 2016, \apjl, 828, L22

\bibitem[{{Shappee} {et~al}\mbox{.}(2018{\natexlab{a}}){Shappee}, {Holoien},
  {Drout}, {Auchettl}, {Stritzinger}, {Kochanek}, {Stanek}, {Shaya}, {Narayan},
  {Brown}, {Bose}, {Bersier}, {Brimacombe}, {Chen}, {Dong}, {Holmbo}, {Katz},
  {Munnoz}, {Mutel}, {Post}, {Prieto}, {Shields}, {Tallon}, {Thompson},
  {Vallely}, {Villanueva}, {Denneau}, {Flewelling}, {Heinze}, {Smith},
  {Stalder}, {Tonry}, {Weiland}, {Barclay}, {Barentsen}, {Cody}, {Dotson},
  {Foerster}, {Garnavich}, {Gully-santiago}, {Hedges}, {Howell}, {Kasen},
  {Margheim}, {Mushotzky}, {Rest}, {Tucker}, {Villar}, {Zenteno}, {Beerman},
  {Bjella}, {Castillo}, {Coughlin}, {Elsaesser}, {Flynn}, {Gangopadhyay},
  {Griest}, {Hanley}, {Kampmeier}, {Kloetzel}, {Kohnert}, {Labonde}, {Larsen},
  {Larson}, {Mccalmont-everton}, {Mcginn}, {Migliorini}, {Moffatt},
  {Muszynski}, {Nystrom}, {Osborne}, {Packard}, {Peterson}, {Redick}, {Reedy},
  {Ross}, {Spencer}, {Steward}, {Van Cleve}, {Cardoso}, {Weschler}, {Wheaton},
  {Bulger}, {Lowe}, {Magnier}, {Schultz}, {Waters}, {Willman}, {Baron}, {Chen},
  {Derkacy}, {Huang}, {Li}, {Li}, {Li}, {Rui}, {Sai}, {Wang}, {Wang}, {Wang},
  {Xiang}, {Zhang}, {Zhang}, {Zhang}, {Zhang}, {Zhang}, {Zhao}, {Brown},
  {Hermes}, {Nordin}, {Points}, {Strampelli}, \& {Zenteno}}]{Shappee18_18bt}
{Shappee} B.~J. {et~al.}, 2018{\natexlab{a}}, ArXiv e-prints

\bibitem[{{Shappee} {et~al}\mbox{.}(2016){Shappee}, {Piro}, {Holoien},
  {Prieto}, {Contreras}, {Itagaki}, {Burns}, {Kochanek}, {Stanek}, {Alper},
  {Basu}, {Beacom}, {Bersier}, {Brimacombe}, {Conseil}, {Danilet}, {Dong},
  {Falco}, {Grupe}, {Hsiao}, {Kiyota}, {Morrell}, {Nicolas}, {Phillips},
  {Pojmanski}, {Simonian}, {Stritzinger}, {Szczygie{\l}}, {Taddia}, {Thompson},
  {Thorstensen}, {Wagner}, \& {Wo{\'z}niak}}]{Shappee16}
{Shappee} B.~J. {et~al.}, 2016, \apj, 826, 144

\bibitem[{{Shappee} {et~al}\mbox{.}(2018{\natexlab{b}}){Shappee}, {Piro},
  {Stanek}, {Patel}, {Margutti}, {Lipunov}, \& {Pogge}}]{Shappee18}
{Shappee} B.~J., {Piro} A.~L., {Stanek} K.~Z., {Patel} S.~G., {Margutti} R.~A.,
  {Lipunov} V.~M., {Pogge} R.~W., 2018{\natexlab{b}}, \apj, 855, 6

\bibitem[{{Shappee} {et~al}\mbox{.}(2014){Shappee}, {Prieto}, {Grupe},
  {Kochanek}, {Stanek}, {De Rosa}, {Mathur}, {Zu}, {Peterson}, {Pogge},
  {Komossa}, {Im}, {Jencson}, {Holoien}, {Basu}, {Beacom}, {Szczygie{\l}},
  {Brimacombe}, {Adams}, {Campillay}, {Choi}, {Contreras}, {Dietrich},
  {Dubberley}, {Elphick}, {Foale}, {Giustini}, {Gonzalez}, {Hawkins}, {Howell},
  {Hsiao}, {Koss}, {Leighly}, {Morrell}, {Mudd}, {Mullins}, {Nugent},
  {Parrent}, {Phillips}, {Pojmanski}, {Rosing}, {Ross}, {Sand}, {Terndrup},
  {Valenti}, {Walker}, \& {Yoon}}]{Shappee14}
{Shappee} B.~J. {et~al.}, 2014, \apj, 788, 48

\bibitem[{{Shappee} {et~al}\mbox{.}(2013){Shappee}, {Stanek}, {Pogge}, \&
  {Garnavich}}]{Shappee13}
{Shappee} B.~J., {Stanek} K.~Z., {Pogge} R.~W., {Garnavich} P.~M., 2013, \apjl,
  762, L5

\bibitem[{{Shen} {et~al}\mbox{.}(2012){Shen}, {Bildsten}, {Kasen}, \&
  {Quataert}}]{Shen12}
{Shen} K.~J., {Bildsten} L., {Kasen} D., {Quataert} E., 2012, \apj, 748, 35

\bibitem[{{Skrutskie} {et~al}\mbox{.}(2006){Skrutskie}, {Cutri}, {Stiening},
  {Weinberg}, {Schneider}, {Carpenter}, {Beichman}, {Capps}, {Chester},
  {Elias}, {Huchra}, {Liebert}, {Lonsdale}, {Monet}, {Price}, {Seitzer},
  {Jarrett}, {Kirkpatrick}, {Gizis}, {Howard}, {Evans}, {Fowler}, {Fullmer},
  {Hurt}, {Light}, {Kopan}, {Marsh}, {McCallon}, {Tam}, {Van Dyk}, \&
  {Wheelock}}]{Skrutskie06}
{Skrutskie} M.~F. {et~al.}, 2006, \aj, 131, 1163

\bibitem[{{Smith} {et~al}\mbox{.}(2012){Smith}, {Nichol}, {Dilday}, {Marriner},
  {Kessler}, {Bassett}, {Cinabro}, {Frieman}, {Garnavich}, {Jha}, {Lampeitl},
  {Sako}, {Schneider}, \& {Sollerman}}]{Smith12}
{Smith} M. {et~al.}, 2012, \apj, 755, 61

\bibitem[{{Speagle} {et~al}\mbox{.}(2014){Speagle}, {Steinhardt}, {Capak}, \&
  {Silverman}}]{Speagle14}
{Speagle} J.~S., {Steinhardt} C.~L., {Capak} P.~L., {Silverman} J.~D., 2014,
  \apjs, 214, 15

\bibitem[{{Sullivan} {et~al}\mbox{.}(2010){Sullivan}, {Conley}, {Howell},
  {Neill}, {Astier}, {Balland}, {Basa}, {Carlberg}, {Fouchez}, {Guy}, {Hardin},
  {Hook}, {Pain}, {Palanque-Delabrouille}, {Perrett}, {Pritchet}, {Regnault},
  {Rich}, {Ruhlmann-Kleider}, {Baumont}, {Hsiao}, {Kronborg}, {Lidman},
  {Perlmutter}, \& {Walker}}]{Sullivan10}
{Sullivan} M. {et~al.}, 2010, \mnras, 406, 782

\bibitem[{{Sullivan} {et~al}\mbox{.}(2006){Sullivan}, {Le Borgne}, {Pritchet},
  {Hodsman}, {Neill}, {Howell}, {Carlberg}, {Astier}, {Aubourg}, {Balam},
  {Basa}, {Conley}, {Fabbro}, {Fouchez}, {Guy}, {Hook}, {Pain},
  {Palanque-Delabrouille}, {Perrett}, {Regnault}, {Rich}, {Taillet}, {Baumont},
  {Bronder}, {Ellis}, {Filiol}, {Lusset}, {Perlmutter}, {Ripoche}, \&
  {Tao}}]{Sullivan06}
{Sullivan} M. {et~al.}, 2006, \apj, 648, 868

\bibitem[{{Thompson}(2011)}]{Thompson11}
{Thompson} T.~A., 2011, \apj, 741, 82

\bibitem[{{Tremonti} {et~al}\mbox{.}(2004){Tremonti}, {Heckman}, {Kauffmann},
  {Brinchmann}, {Charlot}, {White}, {Seibert}, {Peng}, {Schlegel}, {Uomoto},
  {Fukugita}, \& {Brinkmann}}]{Tremonti04}
{Tremonti} C.~A. {et~al.}, 2004, \apj, 613, 898

\bibitem[{{Tutukov} \& {Yungelson}(1979)}]{Tutukov79}
{Tutukov} A.~V., {Yungelson} L.~R., 1979, \actaa, 29, 665

\bibitem[{{Uddin} {et~al}\mbox{.}(2017){Uddin}, {Mould}, {Lidman},
  {Ruhlmann-Kleider}, \& {Zhang}}]{Uddin17}
{Uddin} S.~A., {Mould} J., {Lidman} C., {Ruhlmann-Kleider} V., {Zhang} B.~R.,
  2017, ArXiv e-prints

\bibitem[{{Vallely} {et~al}\mbox{.}(2018){Vallely}, {Prieto}, {Stanek},
  {Kochanek}, {Sukhbold}, {Bersier}, {Brown}, {Chen}, {Dong}, {Falco},
  {Berlind}, {Calkins}, {Koff}, {Kiyota}, {Brimacombe}, {Shappee}, {Holoien},
  {Thompson}, \& {Stritzinger}}]{Vallely18}
{Vallely} P.~J. {et~al.}, 2018, \mnras, 475, 2344

\bibitem[{{Wang} \& {Han}(2012)}]{Wang12}
{Wang} B., {Han} Z., 2012, \nar, 56, 122

\bibitem[{{Webbink}(1984)}]{Webbink84}
{Webbink} R.~F., 1984, \apj, 277, 355

\bibitem[{{Whelan} \& {Iben}(1973)}]{Whelan73}
{Whelan} J., {Iben}, Jr. I., 1973, \apj, 186, 1007

\bibitem[{{Wolf} {et~al}\mbox{.}(2016){Wolf}, {D'Andrea}, {Gupta}, {Sako},
  {Fischer}, {Kessler}, {Jha}, {March}, {Scolnic}, {Fischer}, {Campbell},
  {Nichol}, {Olmstead}, {Richmond}, {Schneider}, \& {Smith}}]{Wolf16}
{Wolf} R.~C. {et~al.}, 2016, \apj, 821, 115

\bibitem[{{Wright} {et~al}\mbox{.}(2010){Wright}, {Eisenhardt}, {Mainzer},
  {Ressler}, {Cutri}, {Jarrett}, {Kirkpatrick}, {Padgett}, {McMillan},
  {Skrutskie}, {Stanford}, {Cohen}, {Walker}, {Mather}, {Leisawitz}, {Gautier},
  {McLean}, {Benford}, {Lonsdale}, {Blain}, {Mendez}, {Irace}, {Duval}, {Liu},
  {Royer}, {Heinrichsen}, {Howard}, {Shannon}, {Kendall}, {Walsh}, {Larsen},
  {Cardon}, {Schick}, {Schwalm}, {Abid}, {Fabinsky}, {Naes}, \&
  {Tsai}}]{Wright10}
{Wright} E.~L. {et~al.}, 2010, \aj, 140, 1868

\end{thebibliography}
